\documentclass[10pt]{article}
\usepackage{amsmath}
\usepackage{amssymb}
\usepackage{graphicx}
\usepackage{graphicx,psfrag,amsmath,amssymb,amsfonts,bbm,latexsym,color,dcolumn,bm, setspace}
\usepackage[includeheadfoot,margin=2cm]{geometry}
\usepackage{color}
\usepackage{soul}
\usepackage{ulem}
\usepackage{subfigure}

\usepackage{cite}
\usepackage{color} 
\usepackage[labelfont=bf,labelsep=period,justification=raggedright]{caption}
\makeatletter
\renewcommand{\@biblabel}[1]{\quad#1.}
\makeatother
\date{}
\begin{document}
% Title must be 150 characters or less
\begin{flushleft}
{\Large
\textbf{Universal Sequence Replication, Reversible Polymerization and Early Functional Biopolymers: A Model for the Initiation of Prebiotic Sequence Evolution}
}
% Insert Author names, affiliations and corresponding author email.
\\
Sara Imari Walker$^{1,2,3,4}$, 
Martha A. Grover$^{1,2}$
Nicholas V. Hud$^{1, 3*}$
\\
\bf{1} NSF/NASA Center for Chemical Evolution, Georgia Institute of Technology, Atlanta, GA, USA
\\
\bf{2} School of Chemical and Biomolecular Engineering, Georgia Institute of Technology, Atlanta, GA, USA
\\
\bf{3} School of Chemistry and Biochemistry, Georgia Institute of Technology, Atlanta, GA, USA
\\
\bf{4} Current address: BEYOND: Center for Fundamental Concepts in Science, Arizona State University, Tempe, AZ, USA
\\
%\bf{*} hud@chemistry.gatech.edu
\end{flushleft}
% Please keep the abstract between 250 and 300 words
\section*{Abstract}
% earliest stage of chemical evolution is difficult b/c it must connect directly to prebiotic chemistry. 
Many models for the origin of life have focused on understanding how evolution can drive the refinement of a preexisting enzyme, such as the evolution of efficient replicase activity. Here we present a model for what was, arguably, an even earlier stage of chemical evolution, when polymer sequence diversity was generated and sustained before, and during, the onset of functional selection. The model includes regular environmental cycles ({\it e.g.} hydration-dehydration cycles) that drive polymers between times of replication and functional activity, which coincide with times of different monomer and polymer diffusivity. Template-directed replication of informational polymers, which takes place during the dehydration stage of each cycle, is considered to be sequence-independent. New sequences are generated by spontaneous polymer formation, and all sequences compete for a finite monomer resource that is recycled via reversible polymerization. Kinetic Monte Carlo simulations demonstrate that this proposed prebiotic scenario provides a robust mechanism for the exploration of sequence space. Introduction of a polymer sequence with monomer synthetase activity illustrates that functional sequences can become established in a preexisting pool of otherwise non-functional sequences. Functional selection does not dominate system dynamics and sequence diversity remains high, permitting the emergence and spread of more than one functional sequence. It is also observed that polymers spontaneously form clusters in simulations where polymers diffuse more slowly than monomers, a feature that is reminiscent of a previous proposal that the earliest stages of life could have been defined by the collective evolution of a system-wide cooperation of polymer aggregates. Overall, the results presented demonstrate the merits of considering plausible prebiotic polymer chemistries and environments that would have allowed for the rapid turnover of monomer resources and for regularly varying monomer/polymer diffusivities.
% Please keep the Author Summary between 150 and 200 words
% Use first person. PLoS ONE authors please skip this step. 
% Author Summary not valid for PLoS ONE submissions. 
%\section*{Author Summary}
\section*{Introduction}
A key question in the origin of life is how the first biopolymers self-organized from simple chemical building blocks into increasingly complex systems with the capacity to cooperate and evolve \cite{SzathmaryMS1997}. Presently, most research that involves experimental models of early chemical evolution and function-based selection has necessarily utilized molecular systems that would have been possible only after considerable evolution of informational polymers \cite{Spiegelman1972,Ellington1990, Tuerk1990, Wochner2011}. Likewise, most theoretical models of early evolution have only considered how evolution could have produced the most efficient self-replicating entity after functional polymers had already emerged \cite{Eigen1971, Eigen1977a, Eigen1988, SSCS, Joyce2002a}. However, proposed scenarios for the chemical origins of life generally include, either explicitly or implicitly, a stage prior to functional polymer evolution when there was an initial buildup of informational polymers with random sequences -- a stage that may have coincided with the selection of the first functional sequences \cite{Joyce2002b, WuHiggs}. Exploring this early stage in chemical evolution is particularly difficult, being so obscured by the ensuing evolutionary history that even the identity of the first replicating material remains a subject of debate, with hypotheses ranging from RNA or some variety of proto-RNA \cite{RNAworld, Nielsen, Joyce2002b, Eschenmoser2007, Engelhart2010}, to peptides \cite{Lee1996}, and even inorganic clay surfaces \cite{CairnsSmith1982}. It also remains challenging to determine the environmental context from which the first informational polymers emerged, as illustrated by the wide variety of proposed sites for the origin of life, including (but not limited to): tidal shores \cite{Miller1995, Com2002}, deserts \cite{HudAnet}, hydrothermal vents \cite{Corliss81}, mineral surfaces \cite{Huber98, Wach, Hazen2010}, the eutectic phase of ice \cite{SM,Monnard2003,Monnard2008}, deep underground \cite{Pedersen2000}, and even atmospheric aerosols \cite{Dobson2000}.

Computer simulations offer a possible means to efficiently explore the potential merits of different scenarios for the emergence of polymer cooperativity and functionality \cite{Szathmary2000, Joyce2002a, SSCS, WuHiggs, Hogeweg2003, MYW, MYZH,  KCS, Hogeweg2009, KC}. Here we explore simulations of polymer sequence and population evolution that incorporate several potentially important physical and chemical concepts of non-enzymatic replication and sequence evolution of prebiotic informational polymers, including: informational polymers with reversible backbone linkages\cite{Lynn1997, Lynn2001, Bean2006, HudLynn2007, Ura2009, Li2011}, environmental cycling \cite{Lahav1978,  HudAnet, ApelDeamer2005, Hazen2009, Fishkis2011}, limited molecular diffusion \cite{Szathmary2000, KCS, Hogeweg2009}, monomer recycling \cite{King82, King86, CN, Engelhart2010}, and template-directed synthesis without sequence restrictions \cite{Deck}. Here we introduce the term Universal Sequence Replication (USR) to represent the possibility that prebiotic template-directed synthesis provided a means for the replication of polymers of a particular chemical structure ({\it i.e.} backbone and side chain structure), regardless of monomer sequence (such that replicative rate constants are, at least to first-order approximation, sequence-independent). USR should be distinguished from the more general framework of non-enzymatic template directed synthesis, which includes chemistries where certain sequences may have a replicative advantage; USR can be seen as a special case where the intrinsic fitness landscape is flat and no sequences possess an intrinsic replicative advantage.

Although the framework presented here shares some individual features with other models, their unification in the model presented provides a fresh perspective on a plausible scenario for environmentally-driven early emergence and evolution of informational polymers. Moreover, we diverge from most previous models in that we address the likely possibility that a nascent population of prebiotic informational polymers would have initially possessed no functional sequences (including no replicase activity). As such, we draw a distinction between the autonomous or mutually dependent ``replicators'' that have been explored in many previous models ({\it e.g.} quasispecies and hypercycle models \cite{Eigen1971, Eigen1977a, Eigen1988}), and the implications of environmentally-driven USR that we explore here. This distinction highlights that the former class of models relies upon the self-replicating capacity of a particular polymer sequence or group of sequences, {\it i.e.} these models assume that functional polymers are already present in the extant population; whereas in the model explored here such a complex function has not yet emerged. Instead, USR posits that all polymers, regardless of monomer sequence, were on equal footing prior to the emergence of functionality. 

Kinetic Monte Carlo simulations were used to explore the dynamics of populations of informational polymers that are formed by spontaneous polymerization, replicated by USR, and subject to hydrolysis in a diffusion-limited environment. The simulations reveal that a population of nonfunctional polymers is evolvable in the sense that new polymer sequences are continually introduced to a diverse extant population, and selection for new functional sequences can occur. The observed dynamics provide insights into a robust mechanism for the early exploration of sequence space, including how functional sequences might become established in an initially random sequence pool. Although USR imposes a flat replication fitness landscape, local feedback resulting from localized resource recycling and limited-diffusivity permit selection of functional sequences. A key result of our simulations is that functional selection does not dominate the system dynamics and species diversity remains high, where we define a species as a population of polymers with identical sequence. High species diversity during functional evolution permits the emergence and spread of more than one functional sequence, where nucleation of functional sequences may be temporally and/or spatially separated.  Furthermore, functional sequences need not become permanently fixed within a population to beneficially impact population level dynamics. This rudimentary form of polymer cooperativity illustrates how evolution may have progressed at a level of the polymer pool before enzyme-based polymer replication or compartmentalization had emerged. The results presented suggest chemical features of candidate polymers and environments that might have initiated functional evolution of informational polymer populations in the origin of life.

\section*{Methods}
In constructing our model, we implemented a relatively small number of parameters to explore select chemical features of candidate prebiotic polymers and their physical setting, as well as the feedback between polymer and monomer populations. We first qualitatively outline the most salient features of the prebiotic scenario captured by our model, and then describe the specific details of how the model is numerically implemented. 

\subsection*{Model Description} \label{MD}
In this section, we briefly describe the physicochemical features of our model, along with the associated adjustable model parameters. All model parameters (apart from enzymatic functional activity) are sequence-independent and are quoted in dimensionless units (see Supporting Information for a detailed description of model parameterization, Methods S1, including a table of parameters and values used in simulations, Table S1). Our primary motivation in choosing {\it all} kinetic rates to be sequence-independent was to focus on the role of the environment ({\it e.g.} cycling, diffusion) in driving system dynamics and functional selection (without biases introduced by differential kinetic rates). One exception is enzymatic functional activity, which is sequence-dependent in our model. 

\paragraph{Regular environmental cycles.} Environmental cycles would have occurred regularly on the prebiotic Earth ({\it e.g.} day-night, tidal, seasonal, hot-cold, freeze-thaw, hydration-dehydration, etc.). In the model presented here we appeal to hydration-dehydration cycles, such as those driven by tidal fluxes or day-night cycling, to provide an energetic source for the assembly of monomers into polymers, where physicochemical properties vary with the environmental phase. Polymerization via spontaneous assembly and USR via template-directed synthesis occur during the hot-dry conditions of the dehydrated phase. Polymer degradation and diffusion of monomers and polymers occur in the cool-wet conditions of the hydrated phase. Additionally, functional polymers (when present in the extant population) only exhibit catalytic activity during the hydrated phase, when cool-wet conditions promote the folding of a polymer into its active state. 

\paragraph{Reversible polymerization.} The model is based upon informational polymers with reversible backbone linkages \cite{Li2011, Ura2009, HudLynn2007, Bean2006, Lynn2001, Lynn1997}. During the hydrated phase, condensation polymers are subject to spontaneous degradation (hydrolysis), governed by the first-order rate constant $k_h$. Monomers liberated via polymer hydrolysis are added back to the local monomer population, creating localized feedback between polymer and monomer concentrations. 

\paragraph{Finite monomer concentration.} The total number of monomeric units ({\it e.g.} nucleotides) in monomer and as polymeric residues is constant in our model system, with equal numbers of the two monomer species labeled $A$ and $B$. The choice to focus on closed mass systems was motivated by our goal of connecting to chemically realistic scenarios, where monomer concentration would have been naturally limited by available resources. For systems with a finite supply of resources, sustainable polymerization requires recycling through turnover of resources via polymer hydrolysis \cite{Engelhart2010, King86, King82, CN}. An exception is the case where new monomers are generated by a functional polymer that acts as a monomer synthetase, where monomer production is still limited by the availability of precursor molecules, which are also a finite resource (see below). 

\paragraph{Surface confinement and limited diffusion.} 
Incomplete mixing, limited diffusion, and surface attachment have been shown to promote template-directed synthesis \cite{Deck, LBvK1998} and to limit the deleterious effects of ``parasites'' (non-functional polymers) on selection of functional sequences \cite{Szathmary2000, KCS, BH}. In the present model, we therefore confine polymer and monomer movement to a surface, as might occur on a mineral with a thin film of water (or a more viscous solution) covering the surface, or other surface-binding phenomenon that permits the reversible association of monomers and polymers and limited movement in two dimensions. The reaction-surface is modeled as a square lattice (see below) where diffusion of monomers and polymers between neighboring sites is governed by the hopping rates $k_m$ and $k_p$, respectively, subject to the physical constraint condition $k_m \geq k_p$. Diffusion only occurs during the hydrated phase of the cycle, when water activity is high. During the dehydrated phase, when the surface is dry, or the viscosity of the thin film is high, diffusion between lattice sites is considered to be negligible, with $k_m = k_p = 0$.

\paragraph{Universal sequence replication (USR).} Recent studies have shown that altering the chemistry of nucleic acid backbone linkages ({\it e.g.} with a reversible linkage) \cite{Li2011, HudLynn2007, Lynn1997, Lynn2001}, or binding of template strands to a surface ({\it e.g.} reduced mobility compared to monomers) \cite{Deck}, allows for accurate template-directed synthesis for a wide range of sequences ({\it i.e.} approaching a form of USR). The effects of idealized USR are explored in our model, {\it i.e.} all polymers have the same intrinsic rate constant for replication, parameterized by the third-order rate constant $k_r$. Despite the fact that all polymers share the same inherent replicative fitness, replication propensity is not the same for each polymer in each replication cycle, as the replication probability for a given polymer depends on both $k_r$ and the local monomer concentrations. Stochastic variation in the spatial distribution of free monomers results in spatiotemporally varying rates of polymer replication, creating a dynamic fitness landscape. As shown below, in the absence of functionality, this selective advantage is random, acting on local populations rather than individuals, with populations sequestering the most resources having an increased chance of survival, independent of their sequence distribution. 

\paragraph{Spontaneous polymer assembly and exploration of sequence space.} 
To simplify our model, we invoke spontaneous polymer assembly as the only means to generate novel sequences ({\it i.e.} we do not consider the effects of mutations during replication or of genetic recombination, which have been extensively studied elsewhere, {\it e.g.} see \cite{Eigen1971, Eigen1988, Szathmary1989, Szathmary2000, Lehman2011, KCS, KC}). In this framework, polymer degradation allows continual exploration of sequence space as novel sequences are introduced via spontaneous assembly of free monomers. The rate of spontaneous assembly (to be contrasted with template-directed assembly) is governed by the second-order rate constant $k_s$, and local monomer concentrations. Throughout this work we set $k_s = 10^{-7}$, corresponding to a maximum global production rate of approximately $1.5$ new sequences per dehydrated phase when all monomers are free (with lower rates when some monomers are sequestered in polymers). This value is intended to reflect a less efficient process of spontaneous polymerization as compared to template-directed synthesis.

\paragraph{Emergence of a functional polymer.} \label{Funcpolymer} A critical stage in the origin of life was the emergence of the first functional informational polymers \cite{HGCS}. We therefore explore with our model the case where a functional sequence is discovered by the random appearance of the sequence. Ma and coworkers have previously argued that in an RNA world, where polymer replication is accomplished without the need for enzymatic polymers ({\it i.e.} by some mechanism of USR \cite{Deck, Horowitz2010, Jain2004}), that a nucleotide synthetase was the first enzyme to emerge \cite{MYZH}. A synthetase would have a clear advantage in an environment where local monomer concentration is the limiting factor for replication. We therefore explore a similar test case. In our simulations, the first polymer to emerge with a beneficial function is referred to as an $A$zyme, a polymer capable of synthesizing $A$ monomers from an additional but also finite resource, proto-$A$ ($pA$). The catalytic activity of the $A$zyme is governed by the second-order rate constant $k_A$ and local $pA$ concentration. The $A$zyme thereby directly couples the functional dynamics to the local environment ({\it e.g.} local monomer abundance). Additionally, the emergence of a synthetase captures some properties of metabolic replicator models\cite{KC, MYZH}, demonstrating a possible pathway between the pre-functional stage of replicating nonfunctional sequences and the subsequent stage of functional sequence optimization.
% Results and Discussion can be combined.

\subsection*{Kinetic Monte Carlo Implementation} \label{Implementation}

The simulations were implemented via a spatially-explicit hybrid kinetic Monte Carlo algorithm \cite{Alfonsi, Chatterjee}. Here we provide a brief summary of the technical details of our implementation. A more in depth discussion of the numerical implementation is provided in the Supporting Information (Methods S1). 

\paragraph{Initial conditions and the reaction surface.}

The reaction surface is modeled on a $64 \times 64$ square lattice.  Each lattice site represents a locally homogenous reaction domain, characterized by a freely interacting community of monomers and polymers ({\it i.e.} a locally well-mixed environment). The lattice size of $4096$ sites was chosen to be small enough for numerical tractability, yet large enough to allow spatial correlations to be observed for the range of kinetic and diffusive parameters under study (such that any spatial organization observed for the parameter ranges explored here is smaller than the lattice dimensions). We impose periodic boundaries to avoid edge effects. The total number of monomeric units ({\it e.g.} nucleotides) in monomer and as polymeric residues is constant, with equal numbers of the two monomer varieties labeled $A$ and $B$. The initial conditions are homogeneous, with all mass in monomer: each of the $4096$ lattice sites is initialized with $60~A$ and $60~B$ monomers, such that $245, 760$ monomers each of species $A$ and $B$ are evenly distributed over the reaction domain at the start of a simulation run. For functional runs in which the $A$zyme appears, each site is also initialized with $60~pA$ monomers, and with $60~pB$ monomers for simulations where the $B$zyme also appears. No polymers are present at the start of a given simulation run (an exception is for functional runs, which start from an already established pool of random sequence polymers at quasi steady-state, see below).

\paragraph{Defining the polymer species pool.} \label{defn:pool}
The number of possible polymer species, $N$, increases exponentially with polymer length: for a polymer of length $L$ and two residue types there are $N = 2^{L}$ possible sequences. We chose to focus our study on polymer populations with a fixed length $R_L$, meant to approximate the dynamics of a population with a mean length of $\bar{L} = R_L$ residues. For the simulations presented here, $R_L = 20$. Our studies of the dynamics with other fixed polymer lengths, including $R_L = 2, 6$ and $10$, yielded qualitatively similar results. Although using shorter length sequences would have enhanced numerical efficiency, we used oligomers of length 20, as nucleic acids of this length are sufficiently long to adopt folded structures and, arguably, large enough to exhibit initial levels of catalytic activity \cite{Yarus}. Additionally, for $R_L = 20$ the sequence space is sufficiently vast that our simulations never explore all possible sequences within the timescales of our simulations. In other words, the sequence space is larger than what our system can dynamically explore in the space and timescales under study; satisfying the minimal requirement of a system with the potential for unlimited heredity and thus open-ended evolvability \cite{SzathmaryMS1997}.

\paragraph{Assigning polymer composition.}\label{polycomp}
Our initial model implementation focused on polymers with fixed length $R_L$ and any possible sequence diversity ({\it i.e.} any possible arrangement of $A$ and $B$ monomer residues could appear). However, sequences with a ratio $A/B = 1$ ({\it e.g.} composed of $10$ $A$ monomers and $10$ $B$ monomers for $R_L = 20$) were an attractor for the dynamics under study, having the greatest access to available resources. Any deviations from the mean distribution, yielding a sequence population dominated by sequences with $A/B \neq 1$, quickly returned to populations dominated by $A/B = 1$ (as an example consider a pure $A$-residue sequence which only has access to half the resources in the reactor pool and would quickly be outcompeted by other sequences containing some $B$ residues which had greater resource availability). Therefore, the simulations were set such that every sequence nucleated contained both $10$ $A$ monomers and $10$ $B$ monomers. This greatly simplified polymer identification since each unique species need only be identified by a single ID or lineage number, $i$ (a dramatic simplification over tracking the specific sequence structure of each unique lineage). This approximation reduced the size of the relevant sequence space from $N = 2^{20} = 1, 048, 576$ to $184,756$ possibilities, but still retained our requirement that the relevant sequence space be much greater than what our dynamics could spatiotemporally explore within a given simulation run (see previous section), and allowed larger and longer statistical samplings. 

\paragraph{Coupling of diffusive and kinetic events to environmental cycles.}
Our numerical implementation of the processes outlined in the previous section explicitly takes into account environmental cycling in order to accurately capture the dynamics of populations of condensation polymers with reversible linkages. The kinetic Monte Carlo implementation is therefore partitioned into two phases: a dehydrated phase, where all lattice sites are diffusively isolated ({\it i.e.} diffusion is turned off); and a hydrated phase, where lattice sites interact diffusively through polymer hopping events and monomer diffusion. Polymer assembly and replication occur only in the dehydrated phase, and polymer degradation occurs only in the hydrated phase.  For simulations exploring the emergence of a functional sequence in the extant population, an additional kinetic process describing enzymatic catalysis is included in the hydrated phase.

\paragraph{The dehydrated phase.}
During the dehydrated phase, each lattice site $x$ on the two-dimensional lattice is diffusively isolated and treated individually with the standard Gillespie algorithm \cite{Gillespie76}. The dehydrated phase supports the kinetic processes of spontaneous assembly and replication. The probabilities of reaction events are weighted by their relative reaction propensities. Spontaneous polymer assembly occurs with (second-order) reaction propensity
\begin{eqnarray} \label{eqn:as}
a_s &=& k_s AB~,
\end{eqnarray}
and USR via template-directed assembly occurs with (third-order) reaction propensity
\begin{eqnarray} \label{eqn:ar}
a_{r_i} &=& k_r AB~N_{i}~.
\end{eqnarray}
Here $A$, $B$, and $N_{i}$ are the number of individual $A$ monomers, $B$ monomers, and polymers of lineage $i$ (the species ID number) at a specific site $x$. The total rate for polymer assembly and USR via template-directed assembly are governed by their respective rate constants, $k_s$ and $k_r$, {\it and} the amount of available monomer resource. This dependence is essential to study how local resource availability affects the dynamics of polymer populations (for example, this feature leads to nontrivial spatial patterning, see Results). Since system dynamics are environmentally driven, a polymer may copy itself {\it at most} once per hydration/dehydration cycle. Therefore, after each template-directed replication event the total number of polymer templates available for further synthesis is decreased by one unit. In other words, we explicitly take into account that the template will not dissociate from the substrate until the next hydrated phase when dilution can drive dissociation. We note that cycling is often invoked as a mechanism for driving strand dissociation \cite{MCN}, however, it is not always explicitly included in model dynamics. We explicitly include cycling, along with its impacts on generational turnover, to explore the potential evolutionary impact on our model prebiotic system. 
%As $A$ and $B$ monomers are incorporated into polymers over the course of the dehydrated phase, the propensities for polymer formation through replication and spontaneous assembly decrease ({\it i.e.} the rates decrease as resources are depleted over the course of each dehydrated phase).

We note that USR via template-directed assembly is treated as a third-order process with a total rate dependent on resource availability through the nucleation term $AB$ (in addition to the rate constant $k_r$ and availability of the template $N_{i}$ - see eq. \ref{eqn:ar}). This relationship is meant to treat dimer formation as the rate limiting step for formation of full-length polymers, {\it i.e.} for the chemistries under investigation here, dimer association with a template is much stronger than monomer association. USR is treated as a one-step process for formation of full-length sequences, since the majority of polymers will form quickly once the first step of dimerization occurs. We therefore consider these approximations to be consistent with the physicochemical processes modeled, with the benefit that they lead to a simplified implementation of the numerical algorithm while still permitting resource dependence in the rates. The choice of kinetics for spontaneous polymer assembly has similar motivation. Dimers are expected to have stronger association to the reaction surface - {\it e.g.} mineral or clay - than monomers, thereby promoting polymerization once dimerization has occurred. The choice to implement resource dependence via a nucleation term $AB$ (rather than $AA$ or $BB$) is consistent with our implementation of a reduced sequence space. 

\paragraph{The hydrated phase.}
During the hydrated phase, monomers and polymers diffuse, and polymers may degrade or, if functionally active, perform catalysis.  Individual lattice sites are diffusively coupled in the hydrated phase, and the dynamics are therefore modeled with a spatially-explicit hybrid kinetic Monte Carlo algorithm \cite{Alfonsi, Chatterjee}. All kinetic events are treated locally, occurring within an individual lattice site which is modeled as a locally homogeneous reaction domain, and only diffusive events occur between sites. A natural partition between rare and common events occurs due to the large separation in the population densities of monomer and polymer observed in our simulations.  The simulations are therefore hybridized such that monomer site hopping is coarse-grained and treated via mass-action kinetics (see Supporting Information for more details, Methods S1). All other events in the hydrated phase are treated stochastically. We have verified that a full stochastic treatment (via a standard spatially-explicit Gillespie algorithm \cite{Bernstein}) reproduces our hybrid results. Against the background of coarse-grained monomer diffusion, the rare polymer events of diffusion and degradation occur. The probabilities of (rare) reaction events are weighted by their relative reaction propensities. Polymer hydrolysis (degradation) occurs with reaction propensity
\begin{eqnarray} \label{eqn:ah}
a_{h_i} &=& k_h N_{i}~,
\end{eqnarray}
and polymer diffusion (hopping between nearest-neighbor lattice sites) occurs with propensity
\begin{eqnarray} \label{eqn:ap}
a_{p_i} &=& k_p N_{i}~. 
\end{eqnarray}
Here hydrolysis is all or none, and degradation is therefore treated as a first-order one-step process which is stochastically determined. This approximation is made based on the implicit assumption that shorter polymer lengths are less stable, as might occur for cases where $20$mers can maintain stable folded conformations whereas shorter length polymers cannot ({\it i.e.} we assume increasingly shorter length polymers become increasingly less stable to hydrolysis).

\paragraph{Data analysis.}

Data presented for quasi-steady state distribution averages, for explorations of both kinetic ({\it i.e.} replication/hydrolysis) and diffusive parameter space, are the combined result of time-averages over $2500$ cycles and ensemble averages over a small statistical sampling of runs (averaged over $5$ and $10$ runs for kinetic and diffusive parameter space exploration, respectively). Small statistical samples are sufficient given the small spread in simulation values and the length of simulations with large time-sampling statistics. The quasi-steady state distribution is defined as the period when the ratio of polymer to monomer achieves an equilibrium value (with fluctuations due to stochastic effects). We used the term ``quasi'' here to indicate that the sequence population is not static, even at steady-state (see Results). Quasi-steady state distributions are calculated starting at $t = 2500$ cycles (typically steady-state is achieved at $ t= 500-1000$ cycles depending on simulation parameters). Time averages were taken from $t = 2500- 5000$ cycles. Data point error bars correspond to sample standard deviation on the mean time-averaged values. In comparing kinetic and diffusive processes for the results presented in this work it is important to note that the simulation dynamics are dependent on the overall rates of processes, which are dependent on both monomer and polymer abundances. The kinetic  rate constants ($k_s$, $k_r$, and $k_h$) and the diffusive hopping rate constants ($k_m$ and $k_p$) are useful in providing measures of the relative strengths of the processes under study, but  must not be confused with the actual rates for the different kinetic and diffusive processes, which are dynamically determined by the ratios of monomer to polymer in a given simulation and their spatial distribution (as defined above, eqs. \ref{eqn:as}, \ref{eqn:ar}, \ref{eqn:ah}, and \ref{eqn:ap}). 

\paragraph{Simulating functionality.}

In simulations including the $A$zyme, which catalyzes formation of $A$ monomers from $pA$ monomers, two additional processes are added to the hydrated phase, diffusion of $pA$ monomers and catalytic conversion of $pA \rightarrow A$. Diffusion of $pA$ monomers, like diffusion of $A$ and $B$ monomers, is treated via mass-action kinetics in the hybridized algorithm. Enzymatic catalysis by the $A$zyme is added to the rare events of the hydrated phase, with the probability of catalysis calculated from the reaction propensity
\begin{eqnarray} \label{eqn:af}
a_{pA} = k_c~ pA~ N_{A \rm{zyme}}~,
\end{eqnarray}
where $pA$ is the number of $pA$ monomers on a local site $x$, $k_c$ is the catalytic rate constant for conversion of $pA \rightarrow A$ in the presence of the $A$zyme, and $N_{A \rm{zyme}}$ is the total number of $A$zymes on the local site $x$. To illustrate the impact of the emergence of a functional sequence, data was saved at $t = 2500$ cycles for all details of a given simulation run. This data provided the initial starting distribution of monomers and polymer species for the functional runs. The choice of starting at $t = 2500$ cycles is somewhat arbitrary given that the systemic dynamics in the quasi-steady state evolution are time-independent, but was chosen to be sufficiently late in the system evolution that a quasi-steady state had been established ({\it i.e.} the ratio of monomer to polymer was relatively constant). To this initial condition, 60 $pA$ monomers were added to each site (to model a previously untapped resource in the environment) and a single polymer representing the $A$zyme sequence, or a nonfunctional sequence, was inserted on the lattice as a spontaneous assembly event. Results were averaged over twenty-five runs, each with a randomly chosen insertion point for the inoculated sequence. Selection of the inoculation site was weighted by the propensities for spontaneous assembly ({\it i.e.} inoculation was not completely random but determined by the resource distribution in the system as done for any other spontaneous assembly event). The simulations were permitted to run until the inoculated sequence (functional or nonfunctional) died out, or until the sequence had survived for $5000$~cycles. Lifetimes were averaged over survival times for the inoculated sequence lineage (defined as the duration of time where at least one individual of the inoculated sequence is still on the lattice) taken over twenty-five simulation runs. Population size averages, the average number of extant species, and exploration rate were averaged over the sequence lifetime. For example, a polymer that lives $5$ cycles is only averaged over $5$ cycles, and therefore yields much higher variance in the data than nonfunctional simulations, which are averaged over entire populations of thousands of sequences, over thousands of cycles. Simulations including a $B$zyme, catalyzing $pB \rightarrow B$, were inoculated in a similar manner with the initial simulation time taken at $t$ = 4000 cycles.

\section*{Results} \label{results}

For each simulation run, the model system was initialized with $60$ $A$ and $60$ $B$ monomers at each of the $4096$ lattice sites, and no polymers. Starting from this homogeneous distribution, we tracked the spontaneous assembly, replication and spatial propagation of informational polymers over several thousand hydration-dehydration cycles. All stochastic events occur locally ({\it i.e.} within or between neighboring lattice sites); however, a global dynamic between local communities emerges due to diffusive contact. Stochasticity is observed to drive polymer population dynamics. In particular, for the parameter ranges investigated here, no system ever achieved a stationary steady-state population of polymers {\it with fixed sequence information}. Instead, the sequences represented in the polymer population continually change with time (Figure 1A). The {\it total population} of polymers is maintained once the system reaches equilibrium (Figure 1B), but the population is temporally varying with respect to the distribution of polymers among existing sequences and due to the appearance of new sequences -- a state of dynamic kinetic stability (DKS) \cite{Pross2003, Pross2005}. Specifically, as individual polymers degrade, monomer recycling allows a quasi-steady state number of polymers to change in sequence distribution by providing resources for replication and spontaneous polymerization. Moreover, stochastic fluctuations in the number of individuals with a given sequence cause species to have a finite lifetime, while some new sequences that appear even after DKS has been achieved are observed to take hold and propagate in the population (Figure 1). We have verified that the quasi-steady states of the DKS observed in our simulations subsist for thousands of environmental cycles once an equilibrium distribution between monomer and polymer is established (running simulations upwards of $> 20,000$ cycles). Stochasticity is also observed to drive dynamic pattern formation in the spatial distribution of polymers. A common feature of our simulations, for a wide range of parameter space, is the spontaneous appearance of polymer clusters. These localized high concentrations of polymers typically begin as a large number of small clusters that then coalesce into fewer, larger clusters over time (Figure 1C; time evolution movies of cluster formation are provided in Supporting Information, Movie S1 and S2). Depending on the parameters of a given simulation, the space between clusters can be essentially devoid of polymers. An individual cluster is typically composed of multiple polymer sequences that mutually benefit from being part of a cluster. As will be shown below, cluster formation correlates with several important system characteristics, such as local monomer concentration, polymer lifetime, sequence diversity and functional sequence propagation.

\subsection*{Exploring Replication and Hydrolysis Rate Parameter Space}
To explore system characteristics as a function of specific model parameters, we measured and compared ensemble averaged data for a number of system metrics collected for simulation runs in which one or two parameters were varied, with the remaining model parameters held constant. For the first set of simulations presented, the polymer replication and hydrolysis rate constants $k_r$ and $k_h$, respectively, were varied while the rate constants for spontaneous polymer assembly and for monomer and polymer diffusion, $k_s$, $k_m$ and $k_p$, respectively, were held at fixed values. Since our aim is to investigate systemic features and evolutionary potential prior to the onset of functionality, no functional sequences exist in the systems presented in this section. As demonstrated by the data shown in Figure 2, the six system metrics of Average Sequence Lifetime, Average Species Population Size, Number of Extant Species, Total Polymer Population, Sequence Exploration Rate, and Average Local Diversity, each change to varying degrees in response to different values for $k_r$ and $k_h$. 

Of the six metrics shown in Figure 2, the dependence of Average Species Lifetime on variations in $k_r$ and $k_h$ is, perhaps, most intuitive. A species is defined as a population of polymers sharing the same unique sequence of $A$ and $B$ monomers, and the species lifetime is the number of contiguous cycles in which one or more copies of that sequence exists in the system. Due to the large number of possible sequences, it is assumed that a particular sequence will spontaneously appear in the system at most once over the timescales of interest. Figure 2A shows that Average Species Lifetime increases with replication rate constant $k_r$, since higher polymer replication rates result in more \textit{copies} of an extant sequence existing in the system. Species lifetime decreases with increases in the hydrolysis rate constant $k_h$, since all polymers have an increasing probability of spontaneous degradation. For the case of $k_h = 1$, a polymer has a $63\%$ chance of degradation during each hydrated phase. Thus, most polymers do not survive into the next cycle, when they would have the opportunity to replicate. Consequently, for $k_h$ = 1 average polymer lifetime is $ <1$ cycle for all values of $k_r$ considered. In the case of no replication, when $k_r$ = 0, only one individual of any given sequence will ever exist in a simulation and therefore the Average Species Lifetime is determined only by the rate of polymer hydrolysis ({\it i.e.} by the average lifetime of a single polymer).

For $k_r \neq 0$ and $k_h < 1$, we observe that the ratio of $k_h/k_r$ largely determines Average Species Lifetime (note: the ratio $k_h/k_r$ represents the ratio of {\it rate constants} and not the ratio of effective reaction rates which are dependent on local monomer and polymer densities). For $k_h/k_r$ = 10, a lifetime $> 10,000$ cycles is observed, while for $k_h/k_r = 1000$, an average lifetime in the range of $100 - 1000$ cycles is observed. For values of $k_h/k_r \ge 10,000$, the Average Species Lifetime drops below $10$ cycles. A distinction can be made in Figure 2A between systems with an Average Species Lifetime $< 10$ cycles and systems with an Average Species Lifetime of $> 100$ cycles. In the former, sequence lineage ({\it i.e.} species) propagation is minimal, whereas in the latter case, propagation of sequence lineages is robust. This result is also illustrated by other system metrics presented below.

Figure 2B shows the Average Species Population Size for extant species after DKS has been reached. Two regimes are clearly visible. In simulations with $k_h/k_r \ge 10,000$, the Average Species Population Size is $ < 2$. As with species lifetime, a high hydrolysis rate will limit the ability for polymers to replicate even for large values of $k_r$, thereby limiting the propagation and copy number of any given species. For values of $k_h/k_r < 100$ (with $k_r > 0$), Average Species Population Size reaches a plateau value that is positively correlated with $k_r$. When $k_r = 0$, the case of no polymer replication, there can only be one copy of each sequence in the system regardless of hydrolysis rate, as is observed in Figure 2B.

Figure 2C provides a view of how the Number of Extant Species, defined as the average number of unique species present at any time after DKS has been reached, varies with $k_r$ and $k_h$. Again, for a sufficiently high hydrolysis rate ({\it i.e.} $k_h$ = 1), polymers generated in a given cycle are likely to degrade before having the opportunity to replicate, thus limiting the total number of polymers present in the system and thereby the Number of Extant Species. For $k_h/k_r \lesssim 1000$, the Number of Extant Species reaches plateau values that decrease with increasing replication rate constant. This relationship is a direct result of competition for a finite supply of resources: larger replication rates lead to greater competition for the limited supply of resources, resulting in higher species extinction rates. In the case of no replication ({\it i.e.} $k_r$ = 0), the Number of Extant Species is equal to the Total Polymer Population Size, {\it i.e.} the steady-state value is determined only by $k_h$, when all other parameters are held constant. 

The Total Population Size of a system, shown in Figure 2D, is defined as the total number of polymers (regardless of sequence) present after DKS has been reached. Plots of this metric illustrate that for $k_h/k_r \le 100$, a plateau value of approximately $21,000$ total polymers is reached. This value corresponds to roughly $85 \%$ of monomers being sequestered in polymers, and $15 \%$ as free monomers. This upper limit on the number of monomers that can be incorporated into polymers is the result of an artificially imposed limitation on the simulation dynamics. Specifically, replication or spontaneous polymer formation cannot take place within a local environment that contains less than the number of monomers necessary to make a full-length polymer ({\it i.e.} twenty monomers, in our simulations with polymer lengths fixed at twenty). The combined systemic features shown in Figures 2B, 2C, and 2D show that for a wide range of hydrolysis and replication rates the same total number of polymers will be present in the system, but the population will be divided between a smaller number of unique sequences (or extant species) as the rate of replication is increased.

The metric of Sequence Exploration Rate is defined as the rate at which new sequences appear in a given system. Because new sequences arise solely by spontaneous formation, this metric provides a measure of the systemic ability to explore sequence space, an absolute necessity if a system is to evolve through the spontaneous appearance of polymers with functional activity. As shown in Figure 2E, for systems where $k_h$ and $k_r$ values give rise to a considerable percentage of free monomers in a state of DKS ({\it i.e.} small total polymer population sizes), the rate of sequence exploration is only limited by the parameter $k_s$, the rate constant for spontaneous polymer formation. As hydrolysis rates are decreased and replication rates increased, the number of free monomers available for spontaneous polymer formation decreases, thereby decreasing the Sequence Exploration Rate. For the case of $k_r$ = 0, where polymers are generated only by spontaneous assembly, Sequence Exploration Rate remains high for a wide range of $k_h$ values, being only limited by the number of monomers made available via polymer hydrolysis. However, the special case of $k_r = 0$ also corresponds to a system in which no sequences propagate through replication. Thus, for a system to evolve through the spontaneous discovery {\it and} propagation of a functional sequence, it is necessary that both the Sequence Exploration Rate be nonzero and that the Average Species Population Size be greater than one. For the system parameters explored here, values of $k_h/k_r$ between $100$ and $1000$ appear to be within a ``sweet spot'' of compromise between sequence exploration rates and the ability for a sequence to take hold in a system through replication.

Average Local Diversity is the final system-level metric shown in Figure 2. In contrast to the previous five metrics, Figure 2F describes the \textit{spatial distribution} of sequences in the system. Sequence diversity was quantified using the Shannon entropy equation \cite{Shannon}, and was calculated locally at each site on the lattice and averaged over the total number of lattice sites (see Supporting Information for additional mathematical details, Methods S1). Average Local Diversity therefore provides a statistical measure of the extent to which multiple sequences coexist on the same lattice site. As such, it provides a measure of diffusive mixing of populations (discussed below) and competition, whereby spatial regions with low local diversity result either from low diffusivity or high rates of local resource competition that result in fewer unique species. Unlike the previous plots, Average Local Diversity is strongly dependent on $k_h$, rather than the ratio $k_h/k_r$. Low hydrolysis rates promote high local diversity, since extinction rates are low with high Average Species Lifetimes.

\subsection*{Exploring Diffusive Parameter Space} \label{diffusion}

We now present simulations designed to explore the effects of monomer and polymer diffusion rates on the system metrics defined above. Specifically, the polymer diffusive hopping rate, $k_p$, was varied within the range $0 \leq k_p \leq 1.0$, and the monomer diffusive hopping rate, $k_m$, was varied within the range $0 \leq k_m \leq 90$. The simulation values investigated are subject to the condition $k_m \ge k_p$, since polymers cannot diffuse faster than monomers. For the simulations presented, the remaining system parameters ({\it e.g.} the kinetic rate constants) were held fixed at values corresponding to intermediate metrics observed for the simulations presented in Figure 2 ({\it i.e.} $k_r = 10^{-4}$; $k_h = 0.1$; $k_s = 10^{-7}$). As in the previous section, our aim in this section is to investigate systemic features and evolutionary potential prior to the onset of functionality. Therefore, no functional sequences are present. Thus, all observed system characteristics presented in this section (including the onset of spatial patterning) arise in the absence of any polymeric catalytic activity ({\it i.e.} without functional sequences).

\paragraph{Spatial patterning.}

Before discussing system metric results, it is instructive to consider the effects of varying $k_m$ and $k_p$ on the spatial distribution of polymers, as these diffusion-dependent distributions are important for understanding results for all other system metrics, as well as being interesting in their own right. As shown in Figure 3, simulations that have completed $3000$ hydration-dehydration cycles exhibit polymer spatial organization that is highly dependent on the diffusivities $k_m$ and $k_p$, and their relative magnitude. All systems shown in Figure 3 had identical initial conditions -- the system was initialized with a uniform distribution of monomers and no polymers at $t = 0$. For monomer diffusive hopping rates $k_m \ge 0.1$, the total number of polymers in the system is fairly constant, with approximately $10,000$ polymers on the lattice. However, large variations in the distribution of polymers are clearly visible. In simulations where monomers and polymers have low diffusivities ({\it e.g.} $k_m = 0.01$ with $k_p = 0.001$), polymer species tend to stay spatially isolated and localized near their nucleation sites. Localized resource recycling sustains these small communities, with small polymer clusters being homogeneously distributed across the simulation lattice. As $k_m$ is increased from $0.01$ to $10$, with $k_p$ fixed at 0.001, polymer cluster size gradually increases until only one or two dominant clusters are observed after $3000$ hydration-dehydration cycles. A similar trend of increasing cluster size is observed for simulations with a $10$-fold greater polymer diffusion rate, ({\it e.g.} $k_p = 0.01$), with larger, more diffuse clusters observed for greater polymer diffusion rates. Further increase in $k_p$ ({\it i.e.} $k_p = 0.1$ and $1.0$ in Figure 3) leads to a loss of defined clusters, or clustering on a scale that is larger than the simulation lattice. In simulations with no polymer movement, $k_p = 0$, new sequences can only be introduced at a grid point by spontaneous polymerization, and clustering is not observed regardless of $k_m$ value. Additional spatial maps are provided in Supporting Information (Figures S1-S5).

The clustering patterns observed in Figure 3 emerge due to the underlying stochasticity of the dynamics. As the first polymers nucleate and replicate, stochastic fluctuations in populations lead to inhomogeneities in polymer population densities. Populations with a slight initial excess of polymer grow by sequestering free monomers that diffuse to their local vicinity. Isolated polymers migrate toward these regions of concentrated resources or go extinct due to resource competition. Thus, clustering emerges as a result of an indirect form of cooperativity between replicating polymers, where early populations with higher polymer densities gain a fitness advantage. Polymers that exist within a cluster enjoy a local recycling dynamic in which the polymers of a cluster act as a reservoir of monomers: fresh monomer resources for replication become available upon polymer hydrolysis, where monomers are sequestered into polymers before these recycled resources have the opportunity to diffuse away from the cluster. This dynamic can occur because the overall rate of polymerization within a cluster is larger than the polymer diffusion rate, thereby resulting in localization of polymer populations (for polymer diffusivities with $k_p \geq 0.01$ in Figure 3 strong clustering is not observed). Between the clusters, polymer density fluctuates near zero. These low population regions act as physical barriers to the transport of information between clusters, leading each aggregate to have a unique sequence population. In contrast, these polymer-depleted regions permit monomer transport, which, through stochastic fluctuations, allow growing clusters to acquire resources from shrinking clusters, even though the polymer clusters may not be in direct contact. The results shown in Figure 3 also illustrate that, for the parameters used in these simulations, polymer population growth is greatly inhibited if monomer diffusion is too slow. In particular, simulations carried out with $k_m = 0.001$ were almost devoid of polymers. 

\paragraph{Diffusive dependence of system metrics.}

In Figure 4 the six system metrics are shown for the set of simulations in which $k_m$ was varied from $0.001$ to $90$, and $k_p$ from $0$ to $1.0$. For nonzero polymer diffusion rates, Average Species Lifetime tends to decrease with monomer diffusion rates of $k_m \ge 0.1$ (Figure 4A). This trend is associated with the observed clustering at higher monomer diffusion rates. High diffusion rates allow for a small number of species to spread and sequester the majority of available resources, which causes high extinction rates for later-nucleating sequences. Thus, a species will, {\it on average}, have a shorter mean lifetime when a smaller number of species are able to dominate the sequestration of resources. The effect of diffusion on species growth is perhaps more easily appreciated by considering two other system metrics: Average Species Population Size and Number of Extant Species. Average Species Population Size (Figure 4B) shows a positive correlation between population size and polymer (and monomer) diffusion rates, with the average population size increasing with increased diffusivity. This result illustrates that increasing polymer and monomer diffusion rates leads to a decrease in the number of species that are able to dominate the acquisition of resources. Likewise, the Number of Extant Species (Figure 4C) shows that the mean number of extant species, like Average Species Lifetime, decreases with increasing polymer and monomer diffusion rates. In the special case of no polymer diffusion ($k_p = 0$), Average Species Lifetime and Number of Extant species increase, for the most part, with increasing monomer diffusion rates. This distinct trend is apparently due to increased monomer diffusion rates allowing for the replication of polymers before spontaneous hydrolysis, but without the ability for sequences to spread and ``colonize" other regions of the surface, which also limits species population size. As noted above, the total number of polymers in a simulation after $3000$ cycles was relatively independent of $k_p$, for $k_m \ge 0.05$ (Figure 4D). 

In contrast to the results presented above for variations in hydrolysis and replication rates, we observe that the Sequence Exploration Rate (Figure 4E) is essentially independent of polymer diffusion rate and only modestly dependent on the monomer diffusion rate for the parameter ranges explored in Figures 3 and 4. Thus, when considering the ability for a system to evolve through sequence exploration, variations in monomer and polymer diffusivities are more likely to affect the capacity for survival of a newly nucleated sequence than the system's capacity to discover new sequences. For example, in the case of no polymer diffusion ($k_p = 0$), a sequence is not able to reap the benefits of spatial expansion, which permits population growth. On the other hand, a sequence that emerges in a system with high polymer mobility will experience strong competition, and may not take hold in the system. The ability for a {\it functional} sequence to overcome these pressures in explored below. Finally, the metric Average Local Diversity shows a unique response to changes in monomer and polymer diffusion rates. Like Average Species Population Size, Average Local Diversity increases with $k_p$. This positive correlation with polymer diffusion illustrates how more rapid polymer movement leads to more complete spatial mixing of polymer species. In contrast, unlike the five other system metrics, Average Local Diversity is apparently independent of $k_m$, an observation that, taken with other system observations, demonstrates how variations in monomer diffusion rates can affect the spatial distribution of polymers without affecting the local spatial distribution of species diversity.

\subsection*{Demonstrating the Emergence of Functionality} \label{Func}

We now address the potential for an individual, catalytically active sequence to become established in a pre-existing pool of nonfunctional polymers. As introduced in The Model section, we chose for our test case the emergence of a polymer sequence (the “$A$zyme”) that catalyzes the production of $A$ monomers from a previously untapped resource of proto-$A$ ($pA$) monomers. For these simulations, the system was initialized with 60 $pA$ monomers at each lattice site (in addition to the 60 $A$ and $B$ monomers), with a single $A$zyme sequence being introduced after the system of nonfunctional polymers had reached a state of DKS. For the results presented here, the $A$zyme was introduced at $t_A = 2500$ cycles. The catalytic rate constant of the $A$zyme was set sufficiently high that enzymatic activity would only be limited by access to $pA$ monomers. This criterion was satisfied by setting the catalytic rate of the $A$zyme such that one enzyme would be able to convert all $pA$ monomers within its lattice site to $A$ monomers within a single hydrated phase. Thus, the observed impact of the $A$zyme sequence on the system does not depend on the catalytic activity of the $A$zyme, but instead on diffusive access to $pA$ monomers, replication of the $A$zyme sequence by USR, and the spread of this sequence by polymer diffusion.

\paragraph{Functional selection.}

Of the four system parameters explored above, the monomer hopping rate, $k_m$, was selected for variation in a series of simulations in which the $A$zyme sequence ``spontaneously'' appears. This parameter was chosen because diffusion of $pA$ monomers into the vicinity of an $A$zyme, as well as the diffusion of newly synthesized $A$ monomers away from a $A$zyme, was expected to affect the ability for an $A$zyme to become established in a pre-existing population. In Figure 5, the Species Lifetime and Population Size of $A$zyme lineages are shown for simulations in which $k_m$ was varied from $0.001$ to $10$. The polymer hopping rate, $k_p$, was fixed at $k_p = 0.01$, and all other parameters were the same as those used in the simulations presented in Figures 3 and 4. The selective advantage of the $A$zyme over nonfunctional sequences is clear in these simulations, with the lifetime of an $A$zyme sequence (shown in black) being $5$ to $60$ times as many cycles as the {\it average} lifetime of nonfunctional polymers (shown in green) in simulations with identical parameters, but without the appearance of an $A$zyme. The population size of the $A$zyme, for all $k_m$ values explored, was approximately two times larger than the average population size of nonfunctional polymers in simulations where no $A$zyme appeared.

It is important to note that Species Lifetime and Population Size for simulations with nonfunctional polymers are weighted towards longer lifetimes and larger population sizes by sequences that appear early in the simulations. Early-time sequences are able to attain relatively large populations before the free monomer concentration begins to limit replication. Therefore, the average lifetimes and populations of early-time sequences are typically much greater in magnitude than those of nonfunctional sequences that emerge later, {\it e.g.} after DKS has been established. Thus, the observed enhanced species lifetime and population size of the late-appearing $A$zyme is even more significant than it first appears relative to the green curve in Figure 5. To illustrate this point, simulations were carried out in which a nonfunctional polymer was introduced at the same cycle time and lattice site as the $A$zyme (without the introduction of the $A$zyme). As shown by the blue curve in Figure 5A, the selective advantage of the $A$zyme sequence is, as expected, more dramatic when compared to the measured lifetimes and population sizes of the late-appearing nonfunctional sequence for all values of $k_m$.

The lifetime of an $A$zyme sequence appears to be less dependent on $k_m$ than the non-functional sequences, being of similar lifetime from $k_m = 0.01$ to $10$. One exception is the considerable increase in $A$zyme lifetime observed in simulations with $k_m = 0.001$. Under these conditions of very slow monomer diffusion, the lifetime of nonfunctional polymer sequences, even those that appear early in a simulation, are too short for any polymer lineages to become established and for a considerable population of any nonfunctional sequence to subsist. Thus, the results shown in Figure 5 also demonstrate that the $A$zyme is able to survive in an environment that cannot sustain nonfunctional polymer populations. In other words, the $A$zyme, by significantly enhancing its own survival, can become the first sustainable extant sequence in a highly dynamic and previously unsustainable environment.

\paragraph{System-level benefits of a functional sequence.}

The population of the $A$zyme is plotted in Figure 6A as a function of time (in units of cycles) for a simulation with $k_m = 10$. A comparison of this plot with those of simulations with only nonfunctional sequences reveals that the population growth of this sequence is faster and to a similar level of the most successful early-appearance polymers ({\it e.g.} Sequence ID 33 in Figure 1A). The effect of $A$zyme appearance on the overall system can be appreciated by comparing the total polymer population after the appearance of the $A$zyme to the steady-state polymer population that is maintained by nonfunctional sequences (Figure 6B). In these simulations, the $A$zyme becomes well established in the population, but its population growth represents only about $10\%$ of the total increase in polymer population. Thus, most of the newly formed $A$ monomers are used to generate new sequences and to replicate existing nonfunctional sequences. While this result might, at first, be considered deleterious for the $A$zyme and a waste of resources on ``parasitic'' nonfunctional sequences, it must be realized that the allocation of resources to other sequences is positive at a system level, as the capacity to search sequence space and to propagate other functional sequences is enhanced.

A comparison of $2D$ spatial maps of polymer and monomer densities for systems with and without the introduction of the $A$zyme sequence further illustrates the effects of a functional $A$zyme on a system of otherwise nonfunctional polymers (Figure 6C). The location of the initial appearance of the $A$zyme sequence is essentially devoid of polymers at $t = 5000$ cycles in the simulation containing only nonfunctional polymers. In contrast, introduction of the single $A$zyme sequence at $t = 2500$ cycles results in a substantial change in the local and global polymer distribution. A cluster grows around the site of $A$zyme introduction, which eventually merges with nearby clusters of nonfunctional sequences. Within this larger cluster the $A$zyme coexists with nonfunctional sequences that benefit from the temporal increase in local free $A$-monomer concentration. Dramatic differences in $A$-monomer and $B$-monomer densities are also observed across the simulation surface (Figure 6C). The $A$ monomer is no longer a limiting factor in polymer production, and eventually increases to higher than pre-$A$zyme levels at every point on the simulation domain. In contrast, the $B$ monomer becomes a more strongly limiting reagent to polymer production, with $B$-monomer levels dropping well below those of the pre-$A$zyme levels across the simulation surface as more resources are consumed by the larger polymer population.

\paragraph{Functional cooperation over time and space.}

Having demonstrated that a single functional sequence can become established within a system of nonfunctional polymers, we next investigated the effects of adding a second functional sequence that acts as a catalyst for the conversion of a proto-$B$ monomer ($pB$) to $B$ monomer. As shown in Figure 6A, in simulations where 60 $pB$ monomers were initially present at each lattice site, the appearance of a single $B$zyme sequence at $t = 4000$ cycles ($1500$ cycles after the appearance of the $A$zyme) quickly results in a burst of $B$zyme population growth. As was observed for the $A$zyme, the growth in $B$zyme population represents only a fraction of the total increase in the global polymer population size (Figure 6B). A plot of polymer density $1000$ cycles after the appearance of the $B$zyme shows the growth of a large cluster with the $B$zyme population at its center. Other polymer clusters on the surface show substantial growth as a result of the appearance of the $B$zyme.

An important feature of the system explored here is that the selective pressure for a functional polymer can be transient in time and space. Each simulation run shows slightly different dynamical evolution after the appearance of a functional polymer. $A$zyme and $B$zyme population plots and 2{\it D} density plots for a second example are provided in the Supporting Information (Figure S6) for a simulation with $k_m = 1.0$ (as compared to $k_m$ = 10 for the simulations shown in Figure 6). For this simulation where only the $A$zyme appears at 2500 cycles (the $B$zyme does not appear), the $A$zyme sequence goes extinct within the next 6000 cycles. In contrast, when the $B$zyme is nucleated near the center of $A$zyme activity, survival of both functional sequences is enhanced. We note that extinction of the $A$zyme, in the absence of $B$zyme appearance, does not terminate system evolution. Figure S6B illustrates that the pool of polymers benefited from the transient activity of the $A$zyme. Furthermore, the $A$zyme had nearly exhausted its benefit to the system at the time of extinction, having had converted nearly all $pA$ to $A$ monomers. However, a stable cluster emerged where the $A$zyme nucleated (Figure S6C), leading to localized enhancement of polymer density and number of extant species. This result demonstrates that the $A$zyme (or any other functional sequence) is not required to live indefinitely in order to have a positive impact on a system undergoing continuous rounds of USR. Moreover, once the $A$-monomer is no longer a limiting reagent, it would be a distinct disadvantage for the $A$zyme population to remain high: for continued system-level evolution, it is more advantageous for the monomers in the $A$zyme sequences to be recycled into polymers with functions that are needed at later times.

\section*{Discussion}
The physical environment and the molecules available on the prebiotic Earth would have placed tremendous constraints on {\it any} mechanism that led to the evolution of informational polymers with functional activity. Among these constraints would have been limited resources and finite polymer stability. With these particular constraints in mind, we used a kinetic Monte Carlo simulation to explore the emergence of functional polymers when only nonfunctional polymers existed that were all replicated with equal probability, regardless of sequence. The model utilizes a minimal set of adjustable parameters in order to explore the effects of those parameters considered most relevant to this putative early stage of prebiotic evolution. The results of our simulations have revealed the possible existence of regions in physiochemical parameter space that could have supported the constant exploration of sequence space, as well as the selection of functional sequences.

Our simulations demonstrate how variations in polymer hydrolysis and replication rates affect the ability for a pool of informational polymers to explore sequence space, even after an equilibrium population of polymers has been established. As expected, systems with faster polymer hydrolysis rates allow more rapid exploration of sequence space, due to more rapid turnover of resources. However, for a system to take advantage of functional sequences that appear spontaneously, the polymer replication rate must be sufficient to counter the deleterious effect of rapid polymer degradation, otherwise information propagation is not sustainable and no polymer lineages become established. Conversely, low polymer degradation rates can cause extant (and nonfunctional) polymers to unproductively retain monomer resources, severely limiting the rate of sequence space exploration. For the model parameters explored here, we have found that a region of compromise exists for a range of replication and hydrolysis rates that allows a nonzero rate of sequence exploration and a nonzero probability that new sequences become established in the system. 

To demonstrate functional sequence selection we focused on the appearance and propagation of monomer synthetases. For this particular functionality, the appearance of the $A$zyme and $B$zyme in the extant population at different points in space and time illustrates a rudimentary form of cooperativity between two functional lineages that may be spatially and/or temporally separated. We argue that such a scenario for functional sequence emergence and early sequence cooperation is more plausible for the earliest stages of prebiotic evolution than scenarios that require the first functional polymer to have been a much more complicated enzyme ({\it i.e.} a processive polymer replicase \cite{Szathmary1989}), or for the diverse members of a set of enzymes to emerge {\it de novo} at the same point in space and time \cite{Kauffman1993}. As the system evolved, there is no reason not to expect that continued system dynamics would eventually permit more complicated catalytic functionalities to be selected and optimized, perhaps even culminating in the eventual appearance of polymerases \cite{WuHiggs2011} and ligases \cite{MYZH10}.

Many origin of life researchers consider polymer compartmentalization, such as nucleic acid encapsulation in lipid vesicles \cite{SD87}, to be a prerequisite for evolution. It is certainly true that there must exist a means for the co-localization of functional polymers with the ``fruits of their labor''. However, limited diffusion (or incomplete mixing) has been shown to provide a possible alternative to encapsulation \cite{SSCS, Hogeweg2003, Hogeweg2009}, at least in the earliest stages of prebiotic evolution. As shown here, system-level metrics, such as Average Species Size and Exploration of Sequence Space, depend on the rates of molecule movement between lattice sites, illustrating that limited movement, representing a realm between stringent compartmentalization and homogenous mixing, could have been beneficial in the early stages of informational polymer growth and evolution. Furthermore, even without employing explicit compartmentalization, nearly all diffusion-limited regimes explored here support stable and diverse populations of extant sequences, and some promote the dynamic emergence of spatial aggregates, which appear even in the absence of any functional activity. In particular, our simulations illustrate how nonfunctional informational polymers can play a positive role by contributing to a local recycling dynamic that sustains extant polymer populations. As one consequence, in all diffusive regimes explored, the unit of selection (or survival) is not the individual polymer, but local populations of polymers dynamically coupled through resource recycling. These populations act both cooperatively as competitive aggregates and individually through single polymer replication/degradation/diffusion events, and through functional selection of active sequences. Thus, the results presented here reinforce previous assertions that physical compartmentalization is not necessary for prebiotic evolution \cite{Hogeweg2009, SSCS}. 

The effects of parasites are not absent in our models. For example, in simulations where $A$zyme and $B$zyme sequences emerge, the majority of monomer resources created by these functional sequences become incorporated into nonfunctional sequences. The observed dynamics are similar to that of the nonfunctional parasites in the model of K\"{o}nny\H{u} and Cz\'{a}r\'{a}n \cite{KC}. However, in contrast to the dynamics observed in their metabolic autonomous replicator model, where parasites are tolerated by functional sequences but play no active role, parasites in the prebiotic scenario presented here are not completely deleterious. When a synthetase first emerges nonfunctional sequences may be beneficial by providing a localized enhancement of resources in an existing polymer cluster that locally retains the new monomer resources (within recyclable polymers) as the synthetase sequence gradually increases in number. At the very least, nonfunctional sequences that take up monomers generated by the synthetase can later provide raw materials for the continued search of sequence space, thereby increasing the survivability and evolvability of the system as a whole.

In conclusion, we have shown that system-level phenomena can emerge from a pool of monomers and replicating polymers that are governed by a small number of meaningful chemical and physical parameters. Moreover, we have shown that in a system of polymers where there is no intrinsic sequence-specific replicative advantage, evolution can still take place. At the rudimentary level -- before the appearance of functional sequences -- environmental cycling, limited diffusivity, and resource limitation leads to the spontaneous self-organization of polymers into spatial aggregates. Even in the idealized case of universal sequence replication, a dynamic fitness landscape spontaneously appears. When functional sequences eventually appear, they can become established amid the nonfunctional polymers, and enhance the ability for other sequences to evolve across space and time. Taken together, these results allow qualitative predictions about the chemistries and environments that would have facilitated prebiotic sequence evolution. Specifically, our model suggests that the optimal conditions for the earliest stage of abiotic chemical evolution, prior to the onset of functional evolution, would {\it not} have been those that promote stringent maintenance of sequence information {\it per se}, which is considered optimum for most autonomous replicator models \cite{Szathmary1989}, including that of Eigen \cite{Eigen1971, Eigen1988}. On the contrary, the optimum conditions for early informational polymer evolution would have allowed the spontaneous appearance of completely new sequences, and for the existence of new sequences to be just long enough for functional sequences to gain a local selective advantage during subsequent cycles of replication. Future investigations of the earliest replicating polymers of life should therefore place more emphasis on polymer chemistries and environments that allow for rapid turnover of resources and time-varying diffusivities for monomers and polymers. Finally, in light of the possibility that compartmentalization appeared after the onset of functional polymer evolution, the results presented here support a model for the early stages of biopolymer evolution that are dramatically different from that governed by a strictly Darwinian process. That is, these early stages could have been defined by the collective evolution of a system-wide cooperation of polymer aggregates. The same general characteristics have been proposed by Woese for the earliest biological systems, but for reasons based on bioinformatics analyses of extant organisms \cite{Woese2002, VWG2006}.

%\section*{Materials and Methods}
% Do NOT remove this, even if you are not including acknowledgments
\section*{Acknowledgments}
We thank Profs. Frank A. L. Anet and Loren Dean Williams for helpful discussions. This work was jointly supported by NSF and the NASA Astrobiology Program, under the NSF Center for Chemical Evolution, CHE-1004570. SIW additionally acknowledges support from the NASA Astrobiology Institute through the NASA Postdoctoral Fellowship Program.
%\section*{References}

% The bibtex filename
\bibliography{ChemEvoRef}
%\begin{figure}[ht]
%\centering
%{
%\includegraphics[width=6in]{Figure1.jpg}
%}

\appendix
\section{Supporting Information: Methods S1}

\subsection{Computational Methods}

Simulations of the spatiotemporal dynamics of diffusion-limited informational polymers with environmentally driven assembly (both spontaneous and template-directed) and degradation were implemented via a hybrid kinetic Monte Carlo algorithm \cite{Alfonsi, Chatterjee}, as outlined here.

\subsubsection{Model Definition}

Our model describing the dynamics of environmentally-driven recycling of informational polymers includes three kinetic and two diffusive processes:
\begin{enumerate}
\item Sequence-independent spontaneous assembly of polymers, with a rate constant $k_s$
\item Template-directed polymer replication via Universal Sequence Replication (USR), with a rate constant $k_r$
\item Sequence-independent polymer hydrolysis, with a rate constant $k_h$
\item Monomer diffusion, with hopping rate $k_m$ (related to monomer diffusivity ${\cal D}_m$ by eq. \ref{eqn:km})
\item Polymer diffusion, with hopping rate $k_p$ (related to polymer diffusivity ${\cal D}_p$ by eq. \ref{eqn:km}).
\end{enumerate}
The kinetic rate constants $k_s$, $k_r$, $k_h$, and diffusive hopping rate constants $k_m$ and $k_p$ are the tunable parameters in our model. Throughout we use italic and lowercase letters to denote dimensionless simulation values and upper-case letters to indicate the physical values of parameters. Since our aim is to study the influence of kinetic and physical parameters on system dynamics, additional model parameters, including total system mass and cycling rate, are held at a fixed values in this work. This simplified model permits us to explore a wide variety of chemistry-environmental couplings ({\it i.e.} a wide range of kinetic and diffusive parameter space) with the tractable set of just five model parameters.

The corresponding dimensionless mass-action kinetic equations for our model system are:
\begin{eqnarray} \label{eqn:kineticG}
\frac{\partial A}{\partial t} &=& {\cal D}_m \nabla^2 A - R_{L _{1/2}} k_s A B - R_{L _{1/2}} k_r A B \sum_{i} X_i +  R_{L _{1/2}} k_h \sum_{i} X_i \\
\frac{\partial B}{\partial t} &=& {\cal D}_m \nabla^2 B  - R_{L _{1/2}} k_s A B- R_{L _{1/2}} k_r A B \sum_{i} X_i +  R_{L _{1/2}} k_h \sum_{i} X_i \\
\frac{\partial X_i}{\partial t} &=& {\cal D}_p \nabla^2 X_i  +  k_s A B  + k_r A B X_i -  k_h X_i \label{eqn:kineticX}
\end{eqnarray}
where $R_{L _{1/2}}$ is half of the polymer length $R_L$ (the motivation for including this factor is outlined below), and the sum over $i$ runs over all extant sequences ({\it i.e.} all unique species in the population at time $t$). The variables $A$, $B$, and $X_i$ correspond to the dimensionless concentrations of $A$ monomer, $B$ monomer, and polymer species with $ID = i$ respectively. This choice of kinetics is intended to approximate spontaneous assembly as a nucleation event, where the potential barrier for nucleation of a new sequence is high but once surmounted the sequence is easily assembled (see Section 2.2.4 in main text for more discussion). Therefore, the rate of spontaneous assembly is governed by the dimensionless second-order rate constant $k_s$ and the local abundances $A(x,y)$ and $B(x,y)$. Likewise, replication is modeled as a third-order process (describing nucleation on a template), governed by a dimensionless sequence-independent third-order rate constant $k_r$, the local monomer concentrations $A(x,y)$ and $B(x,y)$, and the local abundance of polymer species $i$, $X_i$. Hydrolysis is governed by the first-order rate constant $k_h$ and species abundance (see Section 2.2.4 in main text for more discussion). Length dependence is introduced to the kinetic equations through the factor $R_{L _{1/2}}$. The system has a conserved total mass, {\it i.e} we study a closed-mass system. Defining $M$ as the total mass (the total number of monomeric units in monomer and as polymeric residues), we require $M = R_L \sum_i X_i + A + B$.  

The number of possible polymer sequences $N$ increases exponentially with the polymer length $R_L$; we therefore simplify our model by limiting the space of possible sequences through constraining the number of monomer species, the length of polymers, and the possible sequence content of polymers. We implement our model with two monomer species, labeled $A$ and $B$. This choice is motivated by selecting the minimal system that will permit diversity of polymeric sequences.  We consider only polymers with length $R_L = 20$, and each distinct polymeric sequence contains both $10$ $A$-monomers and $10$ $B$-monomers. This simplifies polymer identification: each unique species need only be identified by a single ID number, $i$, where the ID number contains enough information to fully identify the species. Our approximation to the full sequence space greatly simplifies the model while maintaining the essential dynamical features we wish to capture (see discussion in Section in 2.2.2 and 2.2.3 of the main text). It also justifies our implementation of the factor $R_{L _{1/2}}$ included in eqs. \ref{eqn:kineticG} - \ref{eqn:kineticX}, because both monomer species $A$ and $B$ contribute $R_{L_{1/2}}$ residues to a given polymer. Additionally, the fixed polymer length of $R_L = 20$ was chosen to balance our requirements of having sufficient diversity in possible polymer sequences while keeping the dynamics numerically tractable. The number of possible polymers for $R_L = 20$ with a $50:50$, $A:B$ ratio is calculated by the binomial coefficient, $20!/(10! 10!) = 184,756$. This set is large enough such that the simulations presented here only sample a small fraction (less than 5$\%$) of the total number of possible sequences available in the potential sequence pool. The fixed value $R_L = 20$ can be taken to approximate a reaction-diffusion system supporting polymers with a mean-length of $\bar{L} = 20$ residues. These simplifications greatly alleviate the numerical requirements of our simulations, allowing us to run larger statistical samplings of our {\it in silico} experimental runs, while still permitting us to explore the most salient features of system dynamics. We have also explored systems supporting polymerization of shorter sequences, with $R_L < 20$, and observed that the dynamics are qualitatively similar, where the ratio $\frac{N_m}{N_p}$ scales with polymer length for a given set of kinetic parameters ($N_m$ and $N_p$ are the total number of monomers and polymers, respectively).

\subsubsection{Model Implementation}

The reaction surface is modeled on a $64 \times 64$ square lattice.  The lattice size of $4096$ sites was chosen to be small enough for numerical tractability, yet large enough to allow spatial correlations to be observed, for the ranges of kinetic and diffusive parameter under study. The initial conditions are homogeneous, with all mass in monomer: each of the $4096$ lattice sites is initialized with $60~A$ and $60~B$ monomers. For functional runs each site is additionally initialized with $60~pA$ monomers. No polymers are present at the start of a given simulation run (with the exception functional sequence runs which start from an already established pool of random sequence polymers at quasi steady-state, see below). Periodic boundaries were imposed to avoid edge-effects. However, additional simulations performed with closed boundaries revealed similar dynamics to those presented here. For example, in the case of simulation parameters leading to the emergence of localized clusters of high polymer density, we observe polymers preferentially aggregating at the corners of the square grid (not shown). 

Each of the $4096$ lattice sites on the two-dimensional grid represents a locally homogenous reaction domain, characterized by a locally interacting community of monomers and polymers of different sequences. All kinetic events occur locally (on-site). Interaction between lattice sites occurs through diffusive contact. Over the course of system evolution, large separations in polymer and monomer populations are observed, with a typical simulation ratios (at steady-state) of $\sim \frac{N_m}{N_p} = 300$, where $N_m$ and $N_p$ are the total numbers of monomers and polymers, respectively. To increase numerical efficiency, reaction and diffusion events are therefore partitioned such that relatively rare polymer reaction and diffusion events are treated stochastically and much more frequent monomer diffusion events are treated deterministically. This partitioning is implemented with a spatially explicit hybrid kinetic Monte Carlo algorithm \cite{Alfonsi, Chatterjee} (see below). We have verified that the fully stochastic spatially-explicit kinetic Monte Carlo simulation \cite{Bernstein} quantitatively reproduces the same dynamical phenomena (not shown). 

The kinetic equations outlined in eqs. \ref{eqn:kineticG} -- \ref{eqn:kineticX} do not explicitly account for environmental cycling. However, our numerical implementation via kinetic Monte Carlo simulations must explicitly take into account environmental cycling in order to accurately capture the desired dynamics. Our Monte Carlo implementation is therefore partitioned into two phases: a dehydrated phase, where all sites are diffusively isolated ({\it i.e.} diffusion is turned off); and a hydrated phase, where sites interact diffusively through polymer hopping events and monomer diffusion. The three kinetic processes are partitioned between these two phases such that polymer assembly and replication occur in the dehydrated phase, and polymer degradation and enzymatic catalysis (when functional sequences are extant) occur in the hydrated phase. 

\paragraph{The Dehydrated Phase}  

Each lattice site $x$ on the two-dimensional lattice is diffusively isolated and treated individually with the standard Gillespie algorithm \cite{Gillespie76}. The dehydrated phase supports two kinetic processes:
\begin{itemize}
\item Spontaneous assembly, with propensity:
\begin{eqnarray} \label{eq:a_s}
a_s (x) &=& k_s A(x) B(x)\\ \nonumber
\end{eqnarray}
\item Sequence-independent replication via template-directed assembly, with propensity:
\begin{eqnarray} \label{eq:a_r}
a_{r_i}(x) &=& k_r A(x) B(x) N_{i}(x)\\ \nonumber
\end{eqnarray}
\end{itemize}
Here $A(x)$, $B(x)$, and $N_{i}(x)$ are the {\it number} of $A$ monomers, $B$ monomers, and polymer species $i$ at site $x$ (which is not strictly the same as the dimensionless concentrations cited for eqs. \ref{eqn:kineticG} -- \ref{eqn:kineticX} above). The total propensity for polymer formation is 
\begin{eqnarray}
a_{tot}(x) &=& a_s(x) + \sum_i a_{r_i}(x)~,
\end{eqnarray}
where the sum over $i$ runs over all species indigenous to the site $x$. Events are drawn at random. Since the dynamics are environmentally driven, a polymer may copy itself {\it at most} once per hydration/dehydration cycle. Therefore, after each replication event the total number of polymers available for replication is decreased by one unit. As $A$ and $B$ monomers are incorporated into polymers over the course of the dehydrated phase, the propensities for polymer formation through replication and spontaneous assembly decrease. 

For all simulations presented in this work, the rate constant for spontaneous assembly is set to the dimensionless value $k_s = 10^{-7}$. This value corresponds to a maximal global nucleation rate of $1.47$ new sequences per environmental cycle when no polymers are present (calculated by summing $a_s = k_s \times A B$ over all sites for the case where all mass is in monomer, {\it i.e.} $A = B = 60$ monomers per site for all $4096$ lattice sites).  

%Values for the replicative rate constant range from $k_r  = 0$ -- $ 0.001$ for the data shown in Figure 2 in the main text. The rate constant for replication is set at $k_r = 0.0001$ for the presented data exploring diffusive parameter space and functional evolution. Replication is a local process, depending on the on-site populations of monomer and polymer which vary over both space and time. We therefore cannot assign a global replication propensity. However, a rough lower bound for the mean replication time of a single polymer can be estimated using eq. \ref{eq:a_r} with $N_{i} = 1$: assuming the maximal number of monomers per site of $60$ $A-$ monomers, and $60$ $B-$ monomers, yields a lower bound of a minimal mean replication time via USR of $\tau_{R}|_{min} = 0.27$, $2.7$, and $27$-dehydration phases for $k_r = 10^{-3}$, $10^{-4}$, and $10^{-5}$, respectively. These values represent minimal replicative timescales, given that the presence of polymers reduces the number of monomers per site and therefore increases the replicative timescale since some monomers will be sequestered in polymer, resulting in a correspondingly longer mean timescale $\tau_R$ than the values quoted here. However, these values provide a baseline for comparing replicative timescales with mean polymer lifetimes.

\paragraph{The Hydrated Phase} 

Diffusion and hydrolysis occur when the system is hydrated.  In the hydrated-phase the dynamics are modeled with a spatially-explicit hybrid kinetic Monte Carlo algorithm \cite{Alfonsi, Chatterjee}, where polymer diffusive hopping is treated as a stochastic event as in other spatially explicit treatments which incorporate diffusion in the Gillespie algorithm (see {\it e.g.} \cite{Bernstein}). For the system presented here, a natural partition between rare and common events occurs due to the large separation in the population densities of monomer and polymer as discussed above.  Our simulations are therefore hybridized such that monomer site hopping is coarse-grained by introducing the partial differential equation describing mass-action monomer diffusion:
\begin{eqnarray}
\frac{\partial M(x)}{\partial t} & = & {\cal D}_m \nabla^2 M(x)   \label{eqn:diff}
\end{eqnarray}
where ${\cal D}_m$ is the macroscopic monomer dimensionless diffusion rate (see eq. \ref{eqn:D} below for relation to monomer hopping rate cited in the text), $M(x) = A(x)$ or $B(x)$ is the concentration of monomers per unit area, and $\nabla^2$ is the two-dimensional Laplace operator. Eq. \ref{eqn:diff} is evolved forward in time using a standard finite-difference staggered leapfrog algorithm \cite{Press}, with lattice spacing $dx = 0.2$ and time-step $dt = 0.01$. Implementation of this algorithm is subject to the stability criteria imposed by the Courant Condition yielding numeric stability for ${\cal D}_m < 1$, constraining our simulations to values $k_m < 100$ (see eq. \ref{eqn:D}), for the values of $dx$ and $dt$ implemented in the simulations.

Against the background of monomer diffusion, the rare polymer events of diffusion and degradation occur. These rare processes are: 
\begin{itemize}
\item Polymer hydrolysis (degradation) with propensity
\begin{eqnarray}
a_{h_i} (x) &=& k_h N_i(x)\\ \nonumber
\end{eqnarray}
\item Polymer diffusion (hopping between nearest-neighbor lattice sites) with propensity
\begin{eqnarray} \label{eq:a_p}
a_{p_i}(x) &=& k_p N_{i}(x)\\ \nonumber
 \nonumber
\end{eqnarray}
\end{itemize}
The time-step between rare-events is calculated by determining the value of $\tau$ such that
\begin{eqnarray}
\int_t^{t+\tau} a_{rare}(t') dt' = \ln \left( \frac{1}{\xi} \right) \label{eqn:int}
\end{eqnarray}
 where $a_{rare}$ is the total propensity for rare reactions, 
 \begin{eqnarray} 
 a_{rare} = \sum_{x} \sum_i a_{p_i}(x)  + a_{h_i} (x)~,
\end{eqnarray} 
with events selected globally, and $\xi$ is a random number drawn from the set $(0,1]$ \cite{Alfonsi}.  To calculate the integral in eq. \ref{eqn:int}, the numbers of all rare and common molecules must be followed in time. The deterministic PDE of eq. \ref{eqn:diff} is used to track the evolution of the monomers species $A$ and $B$. The PDEs are integrated forward in time using the staggered leapfrog algorithm until eq. \ref{eqn:int} is satisfied. One stochastic event is then chosen to occur at random from the global pool of possible rare events on the lattice by a standard Gillespie algorithm \cite{Gillespie76}, which is calculated only for rare events. 

\paragraph{Functional Runs}

In simulations including a functional $A$zyme, which catalyzes formation of $A$ monomers from $pA$ monomers, two additional processes are added to the hydrated phase: diffusion of $pA$ monomers, and catalytic conversion of $pA \rightarrow A$. Diffusion of $pA$ monomers is treated the same as $A$-- and $B$-- monomer diffusion using the mass-action monomer diffusion eq. \ref{eqn:diff}. Catalysis is added to the rare events of the hydrated phase, with the probability of catalysis calculated from the reaction propensity:
\begin{eqnarray}
a_{pA} (x) &=& k_c pA(x) N_{j}(x)
\end{eqnarray}
where $pA(x)$ is the dimensionless $pA$ concentration ({\it i.e.} the number of $pA$ monomers at site $x$), $k_c$ is the microscopic catalysis rate for conversion of $pA \rightarrow A$ in the presence of the $A$zyme, and $j$ is the ID number of the $A$zyme. The total propensity for rare events with an extant functional $A$zyme is therefore modified such that $a_{rare} = \sum_{x} (a_{A} (x)  + \sum_i (a_{p_i}(x)  + a_{h_i} (x))$. 

As discussed in the main text, to illustrate the impact of the emergence of a functional sequence, data was saved at $t = 2500$ cycles for all details of a simulation, before continuing to run the simulation with no functional sequences present. This data provided the initial starting distribution of monomers and polymer species for the functional runs. The choice of starting at $t = 2500$ cycles is somewhat arbitrary, but was chosen to be sufficiently late in the system evolution that a quasi-steady state had been established ({\it i.e.} the ratio of monomer to polymer was relatively constant). To this initial condition, 60 $pA$ monomers were added to each site (to model a previously untapped resource in the environment). A single polymer sequence representing the $A$zyme was randomly inserted on the lattice as a spontaneous assembly event (with the choice of site weighted by the propensities in eq. \ref{eq:a_s} for each run). Catalytic efficiency, $k_c$, was set to 100. Results were averaged over twenty-five runs (each with a randomly chosen insertion point for the inoculated sequence). The simulations were permitted to run until the inoculated sequence died out, or until the sequence had survived for $5000$~cycles.   The $B$zyme was later inserted at $t$ = 4000 cycles, catalyzing $pB \rightarrow B$.

\subsection{Dimensionalization of Model Parameters} \label{dim}

Throughout this work, we cite dimensionless microscopic kinetic and diffusive rates used in the kinetic Monte Carlo simulations. To recover dimensioned values we define the dimensionless time and space variables 
\begin{eqnarray}
t = \frac{T}{\gamma}~,~~~~~~~~ x = \frac{X}{\lambda} ~, \label{eqn:scaling}
\end{eqnarray}
where $\gamma$ is a typical time scale, and $\lambda = \sqrt{ D \gamma}$ is a typical spatial scale, and  $D$ is a dimensioned diffusion rate which we scale out of the equations such that:
\begin{eqnarray}
{\cal D}_m =  \frac{D_m}{D}~,~~~~~~~~ {\cal D}_p =  \frac{D_p}{D} \label{eqn:D}
\end{eqnarray}
The system is subject to the physical constraint $\frac{D_p}{D_m} \geq 1$. 
With this parameterization, the dimensionless kinetic rate constants implemented in our simulations (indicated by lower-case letters) are related to their physical counterparts (denoted by upper-case letters) as:
\begin{eqnarray} \label{eqn:micro_rates}
k_h = \gamma K_h~~,~~ k_s = \frac{\gamma K_s}{\lambda^2}~~,~~k_A = \frac{\gamma K_A}{\lambda^2}~~,~~ k_r = \frac{\gamma K_r}{\lambda^4} \label{eqn:rates}
\end{eqnarray}
for hydrolysis (first-order process), spontaneous assembly and enzymatic catalysis (second-order processes), and templated-directed replication (third-order process), respectively. 
Dimensionless concentrations may be made dimensional by the transformation:
\begin{eqnarray}
A = [A] \lambda^2 ~,~~~~~~~~ B = [B] \lambda^2 ~,~~~~~~~~ N_i = [X_i] \lambda^2 ~ \label{eqn:conc}
\end{eqnarray}
where $[A]$ is the concentration of $A$ monomers, $[B]$ is the concentration of $B$ monomers, and $[X_i]$ is the concentration of polymer species with ID number $i$.  The notation $[\ldots]$ is used to indicate concentration in number of molecules per unit area such that the variables $A$, $B$, and $N_i$ correspond to the number of molecules within a spatial region of size $\lambda^2$. As shown below, it is convenient to take the length scale $\lambda$ to be the physical size of a lattice site.

To model diffusion events on the lattice, a relation between the mass-action diffusivities defined above, and the microscopic hopping rates must be defined. The microscopic diffusive hopping rate for polymers is defined as:
\begin{eqnarray} \label{eqn:kp}
k_p = \frac{4 {\cal D}_p}{dx^2}
\end{eqnarray}
where we have used the mean-square displacement for particles subject to Brownian diffusion $\langle r^2 \rangle = 2 d {\cal D}_p \tau$, with dimensionality $d=2$, and a mean displacement $\langle r^2 \rangle = dx^2$, noting that $\tau = k_p^{-1}$ is the hopping time between sites. It is the dimensionless hopping rate $k_p$ which we use as a simulation parameter and therefore cite throughout this work. The dimensional diffusivity $D_p$ can be recovered using eqs. \ref{eqn:D} and \ref{eqn:kp}.

Since monomer diffusion is treated deterministically via a mass-action diffusion equation, the monomer diffusion parameter used in our simulations is ${\cal D}_m$. However, we additionally define a dimensionless monomer hopping rate:
\begin{eqnarray}\label{eqn:km}
k_m = \frac{4 {\cal D}_m}{dx^2}
\end{eqnarray}
which is calculated in an analogous manner to $k_p$. Throughout this work we cite the dimensionless hopping rate $k_m$, rather than the simulation parameter ${\cal D}_m$, to provide an more direct comparison between monomer and polymer diffusion rates and timescales. %The microscopic reaction rate constants must be dimensionless \cite{Gillespie76, Chatterjee}. Since eqs. \ref{eqn:kineticG} - \ref{eqn:kineticX} are already nondimensionlized, the dimensionless microscopic kinetic rate constants are defined relative to physical constants by eq. \ref{eqn:micro_rates}. The parameters $k_h$, $k_s$, and $k_r$ are the parameters used in our simulations and are cited throughout this work.

For numerical simplicity, we normalize the physical size of a lattice site $L$, to $L^2 = 1~ l^2$, where $l$ is a unit of length ({\it e.g.} $l = \mu$m, mm, or cm), such that the number of molecules on a lattice site is equal to its dimensionless concentration ({\it i.e.} $N_i = X_i$ where $X_i$ is the dimensionless concentration of species $i$ cited in eqs. \ref{eqn:kineticG} -- \ref{eqn:kineticX}). Consequently, the non-dimensional size of a lattice site is $dx = 0.2$ for our simulations, and using eq. \ref{eqn:scaling} this yields a constraint condition
\begin{eqnarray}
\sqrt{ D \gamma}  = 5~l, \label{eqn:constraint}
\end{eqnarray}
(where we have used $\lambda = \sqrt{ D \gamma}  = \frac{L}{dx}$), in our model parameterization. Scaling parameters $D$ and $\gamma$ must be chosen to satisfy eq. \ref{eqn:constraint}, to be consistent with our numerical implementation. Because it is easiest to define a timescale $\gamma$ relative to the typical physical length of an environmental cycle, this usually translates to a constraint on the scaling parameter $D$, which is otherwise an arbitrary parameter. This choice therefore does not affect interpretation of our results, and enables straightforward recovery of dimensionalized values for simulation parameters by defining typical length and timescales $\lambda$ and $\gamma$ only.

\subsection{Data Analysis}

In the absence of functionality, the system comes to a quasi steady-state of dynamic kinetic equilibrium after several hundred cycles. The exact timescale depends on the particular diffusive and kinetic rates of a given system. Equilibrium is defined by a quasi steady-state ratio of monomer to polymer. That is, $\frac{d(N_m/N_p)}{dt} \simeq 0$, where $N_m$ is the total number of monomers and $N_p$ is the total number of polymers, with stochastic fluctuations about the equilibrium value. The quasi steady-state distribution is dynamic, with the population of extant polymers continuously changing with time. Nonetheless, certain global measurements, including the total number of polymers, the global and averaged local diversity (defined below), the average number of extant species, and the average extant species population size and lifetime, reach values during the quasi steady-state that are characteristic of a given set of system parameters ({\it i.e.} these values are deterministic, with stochastic fluctuations about a characteristic value). To make comparisons between different systems, with different sets of kinetic and diffusive parameters, we time-averaged these characteristic values during the quasi-steady state system evolution. Data is time-averaged over $2500$ cycles and then ensemble averaged over a small statistical sampling of runs (for work presented here these sample sizes are $5$ or $10$ runs). Small statistical samples are sufficient given the small spread in simulation values and the length of simulations with time-sampling over $2500$ cycles. The quasi-steady state distribution values are calculated starting at $t = 2500$ cycles, to ensure that all systems have achieved a steady-state distribution of monomer to polymer for all parameter ranges explored (typically steady-state is achieved at $ t= 500-1000$ cycles depending on simulation parameters). Error bars correspond to sample standard deviation on the mean time-averaged values. 

For runs illustrating the emergence of a functional sequence, where a random sequence is inoculated at a randomized location on the lattice (weighted by the propensities for spontaneous assembly at each site), lifetimes are averaged over survival times for the inoculated sequence lineage taken over all twenty-five experimental runs. Population size averages, as well as the average number of extant species and exploration rate, are averaged over the sequence lifetime ({\it i.e.} for a polymer that lives $5$ cycles this corresponds to an average over $5$ cycles) and therefore yield much higher variances in the data set than for the nonfunctional simulations which are averaged over entire populations of thousands of sequences, over thousands of environmental cycles. 

\subsubsection{Local Diversity}

Each lattice site is characterized by its sequence population. To calculate the local (on-site) diversity of polymer populations we use the Shannon equation for information \cite{Shannon}, used in studies of species diversity (see {\it e.g.} \cite{Filotas}). Local diversity is calculated for each site individually as,
\begin{eqnarray}
S_L(x,t) = - \sum_i \frac{N_{i} (x,t)}{N_{tot}(x, t)}\ln \left( \frac{N_{i} (x,t)}{N_{tot}(x,t)}\right)
\end{eqnarray}
where $i$ sums over all unique polymer species on the local site $x$, $N_{i}(x, t)$ denotes the local population size of each unique polymer species $i$ on site $x$ at time $t$, and $N_{tot}$ is the total number of polymers on the local site  $x$ at time $t$ ({\it i.e.} $N_{tot}(x,t) = \sum_i N_{i} (x,t)$ ). Local diversity measures the statistical diversity of each lattice site and is spatially averaged over all sites $x$ to yield the average local diversity
\begin{eqnarray}
\langle S_L (t) \rangle = \frac{ \sum_{x} S_L(x,t) }{4096}
\end{eqnarray}
where $4096$ is the total number of local populations (lattice sites) on our $64 \times 64$ square lattice. It is this spatially-averaged local diversity that is reported in our ensemble averaged data analysis (with spatial maps of local diversity included for simulation snapshots).  Average local diversity therefore provides a statistical measure of the extent to which multiple sequences coexist on the same lattice site.  As such, it provides a measure of diffusive mixing of populations and competition, whereby spatial regions with low local diversity result either from low diffusivity or high rates of local resource competition that result in fewer unique species. 

\subsubsection{Similarity Index}

As an added measure of the spatial diversity of the polymer population we utilize a generalization of the of the Jaccard similarity index used in population ecology \cite{Filotas, Chao}. This index provides a statistical measure of the similarity between next-nearest neighbor ``communities'' or lattice sites. Adopting this concept to chemical evolution, the similarity between two lattice sites $\alpha$ and $\beta$ containing $N_\alpha$ and $N_\beta$ unique sequences, respectively, and sharing $N_{\alpha \beta}$ mutual sequences is given by:
\begin{eqnarray}
I_{ \alpha \beta} (t)= \frac{R_\alpha(t) R_\beta(t)}{R_\alpha(t) + R_\beta(t) - R_\alpha(t) R_\beta(t)}
\end{eqnarray}
where $R_\alpha$ is the sum of the relative abundances of the shared species at site $\alpha$:
\begin{eqnarray}
R_\alpha(t) \equiv \sum_{i}^{N_{\alpha \beta}} \frac{N_i (\alpha, t)_{shared}} {N_{tot} (\alpha, t)}~.
\end{eqnarray}
Here $N_i(\alpha, t)_{shared}$ is  the abundance of the $i$-th species shared between sites $\alpha$ and $\beta$ that are on site $\alpha$ at time t, and $N_{tot}$ is the sum total of all polymers of both shared and unshared species at site $\alpha$ ({\it i.e.} $N_{tot}(\alpha, t) = \sum_i N_i (\alpha, t)$ ). An equivalent definitions for $R_\beta$ is obtained with the substitution $\alpha \leftrightarrow \beta$.

A map of similarity of nearest-neighbor sites is produced by calculating the pairwise similarity index $I_{\alpha j}$ at the site $\alpha$, where the index $j$ runs over the eight nearest neighbor communities $j = 1, 2, . . . 8$. For each site, the indices $I_{\alpha j}$ are averaged to yield the site specific value:
\begin{eqnarray}
 I (x_\alpha, y_\alpha)  = \frac{\sum_j I_{\alpha j}}{8} ~~~~~~~~~ j = 1, 2, . . . 8
\end{eqnarray}
The value $ I (x_\alpha, y_\alpha) $ is recorded as the similarity index for site $\alpha$ in the similarity map. This procedure provides a method of identifying regions in the system occupied by highly similar populations of polymers as well as aiding in visual identification of clustered regions as shown in the Supporting Figures below. 

\subsubsection{Sequence Space Exploration Rate, Extant Species, and Species Lifetime}

The total number of species introduced to the system, $N_S (t)$, is tracked over the entire course of each simulation and provides a measure of the total size of sequence space explored as a function of time.  The rate of exploration of sequence space $R_S(t)$, {\it i.e.} the species production rate,  is therefore calculated as the slope of the time-dependent total species, {\it e.g.} $R_S(t) = \dot{N}_S(t)$, discretized as $R_S(t_i) = \frac{N_S(t_i) - N_S(t_{i-1})}{2}$, which achieves a constant (within small fluctuations) and nonzero value during the quasi steady-state system evolution. $R_S(t)$ is time-averaged to obtain a mean exploration rate for each simulation run. Since the total number of extant species, $N_{ES}$ (a directly measured value) achieves a constant value (within small fluctuations) during the quasi steady-state evolution, the number of species created per unit time is equal to the number destroyed as the extant population explores sequence space. An estimate of the mean species lifetime $\tau_S$ is therefore calculated as $\tau_S = N_{ES}/R_S$. 

\section{Figures and Supporting Figures and Legends}

\begin{figure}[ht] \label{fig:kinetics}
\centering{\includegraphics[width=6in]{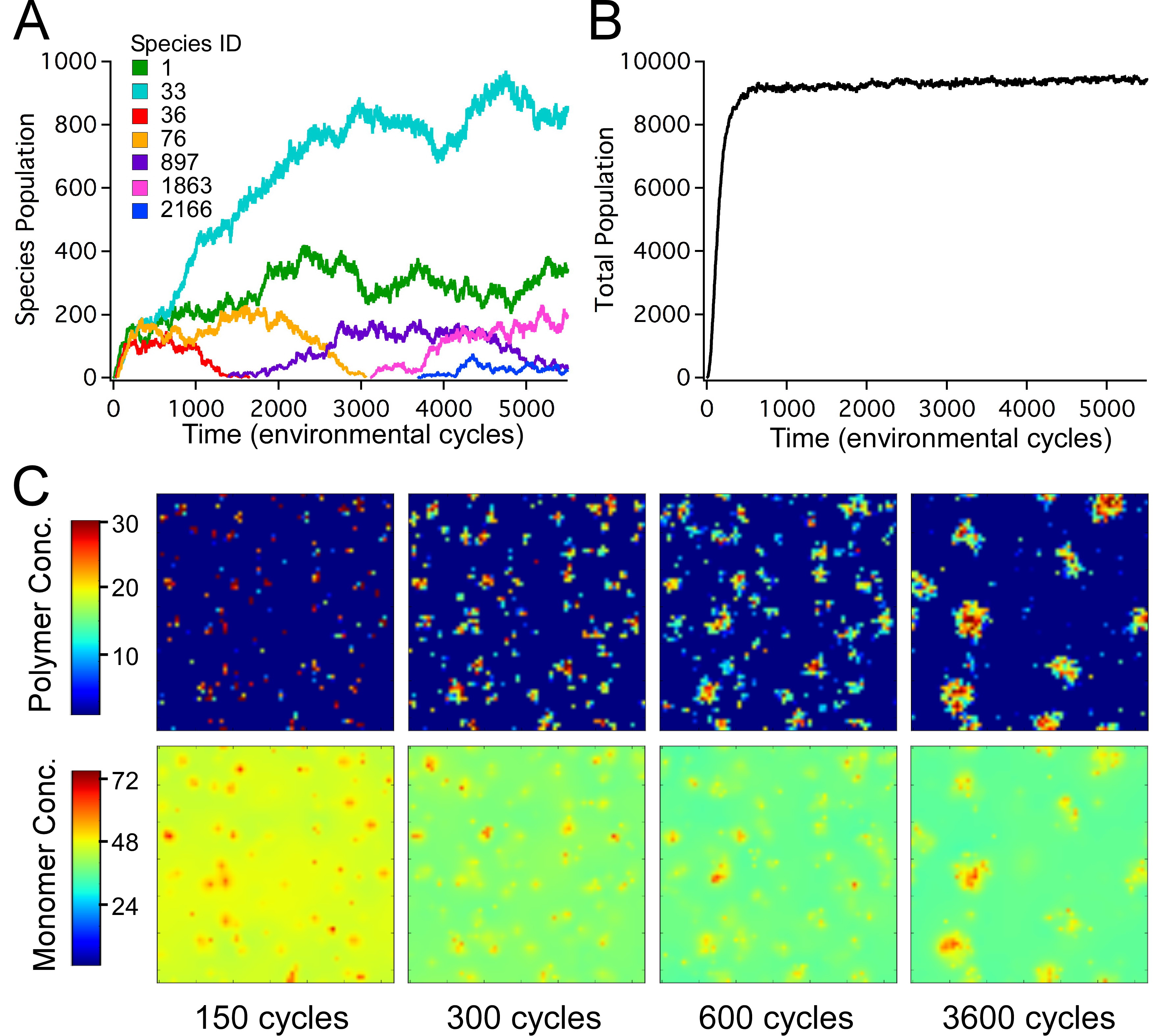}

}
\caption[Optional caption for list of figures]{
{\bf Sequence evolution of the polymer pool.} A: Time evolution of the populations of seven specific sequences; B: Time evolution of the total polymer population; C: Spatial snapshots of the total polymer and monomer concentrations at four representative times. Species ID indicates the order of appearance of the first individual of a particular sequence in the polymer pool. The kinetic rate constants are $k_s = 10^{-7}$, $k_r = 10^{-4}$, and $k_h = 0.1$, and polymer and monomer diffusivities are set as $k_p = 0.001$ and $k_m = 10.0$ sites/cycle, respectively. Units of time are in number of cycles.}
\end{figure}

\begin{figure}[ht]
\centering{\includegraphics[width=6in]{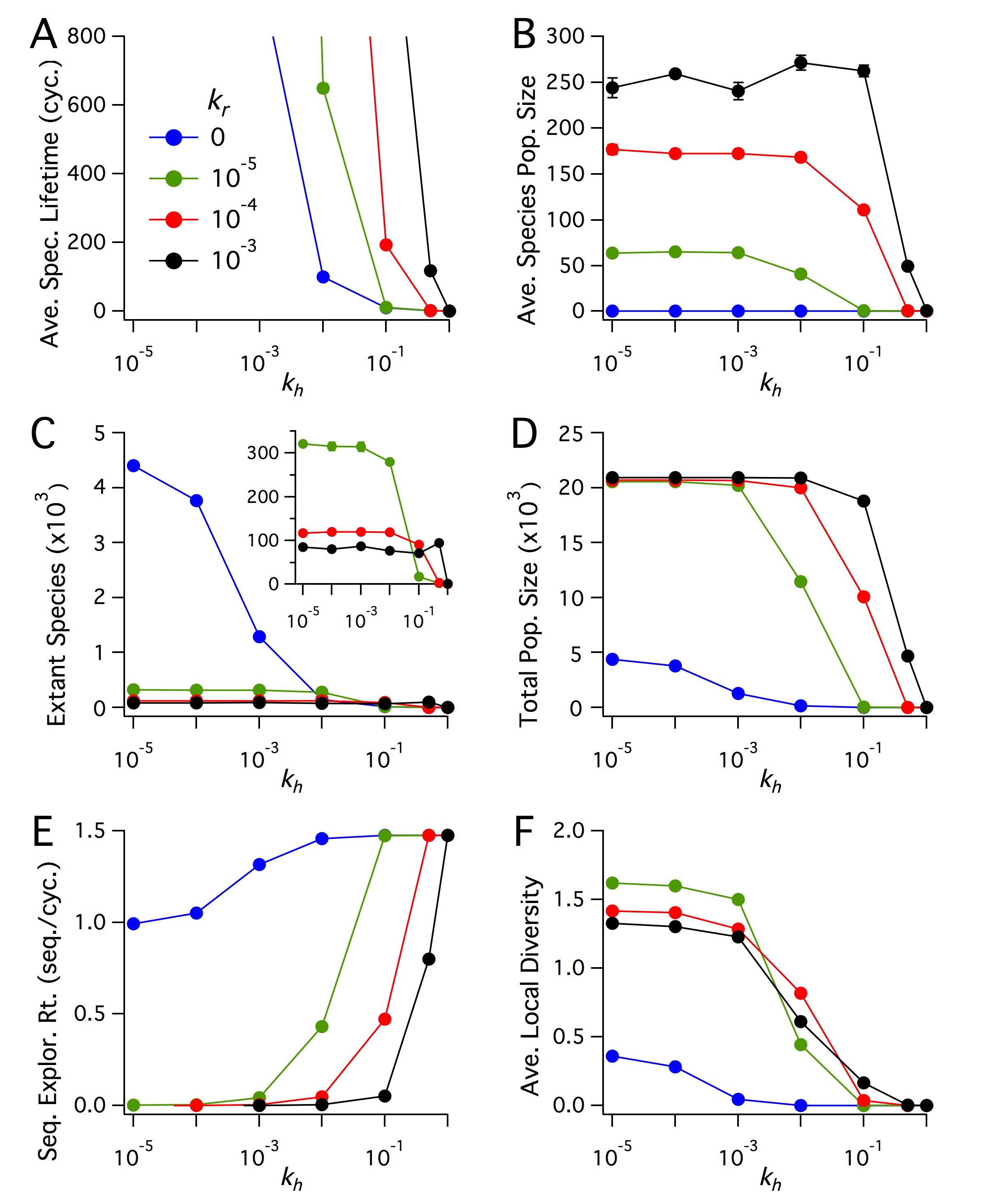}
}
\caption[Optional caption for list of figures]{
{\bf Exploring kinetic parameter space.} Plots of quasi steady-state values of Average Species Lifetime (in units of number of cycles), Average Species Population, Extant Species, Total Population Size, Sequence Exploration Rate (in units of number of novel sequences generated per cycle), and Average Local Diversity. Time averages were taken from $t = 2500- 5000$ cycles, and each point is the ensemble average over five realizations. Error bars denote the sample standard deviation (most are smaller than symbols). The rate constant for spontaneous sequence nucleation is $k_s = 10^{-7}$, and the diffusion rate constants are $k_p = 0.01$ sites/cycle and $k_m = 1.0$ sites/cycle. The blue data set shows the reference case with $k_r = 0$, where no polymers replicate. }
\label{fig:H}
\end{figure}

\begin{figure}[ht]
\centering{\includegraphics[width=6in]{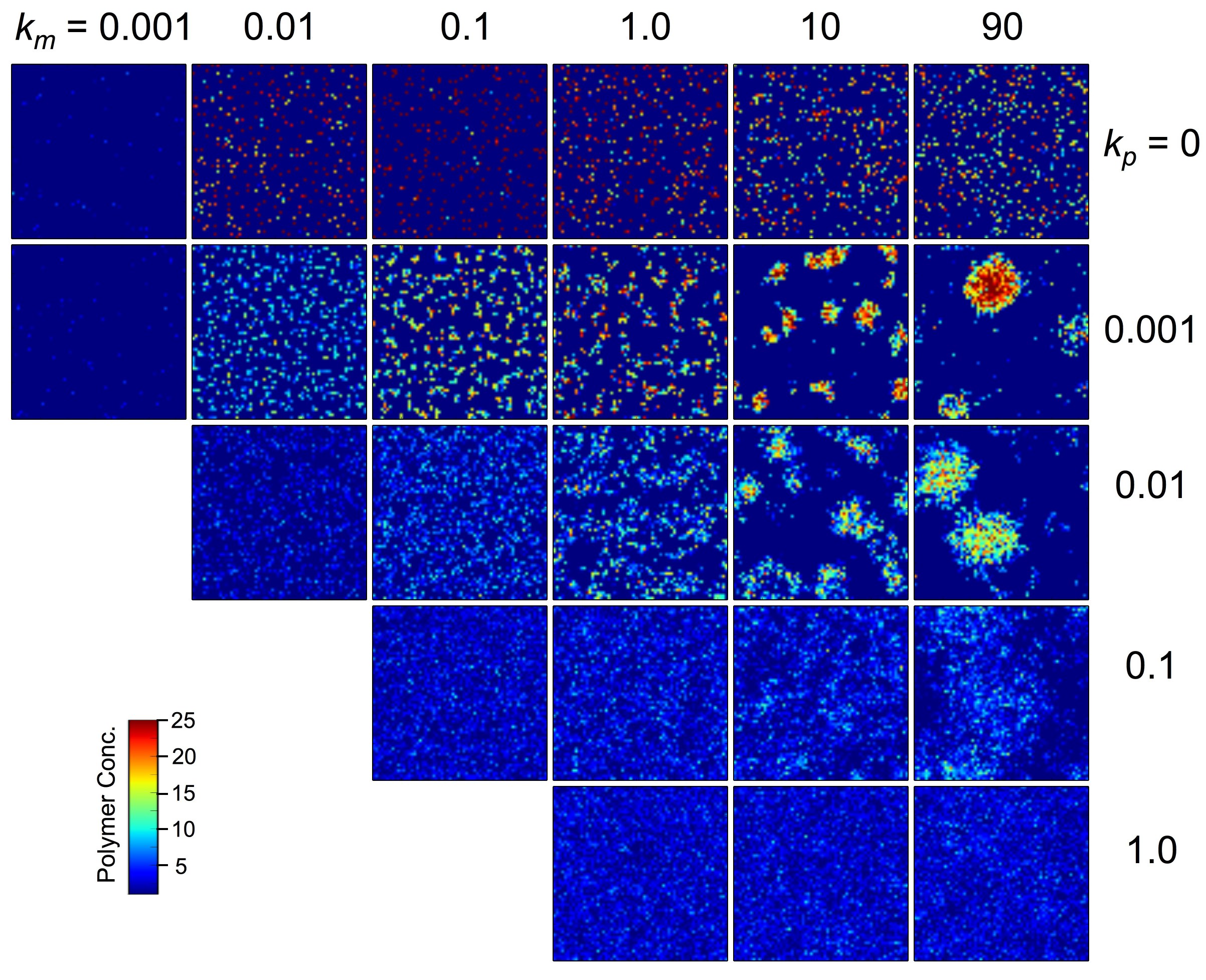}
}
\caption[Optional caption for list of figures]{
{\bf Spatial maps of polymer density.} Each column of images corresponds to simulations run with a different value for monomer diffusivity $k_m$ (in sites/cycle), and each row has a different value for polymer diffusivity $k_p$ (in sites/cycle). All data shown are for kinetic rate constants $k_s = 10^{-7}$, $k_r = 10^{-4}$, and $k_h = 0.1$. All maps correspond to $t = 3000$ cycles. The color scale is in units of polymers/site. Simulations were only run for cases in which monomer diffusivity is greater than or equal to the polymer diffusivity.} \label{fig:spatial_map}
\end{figure}

\begin{figure}[ht]
\centering
{
\includegraphics[width=6in]{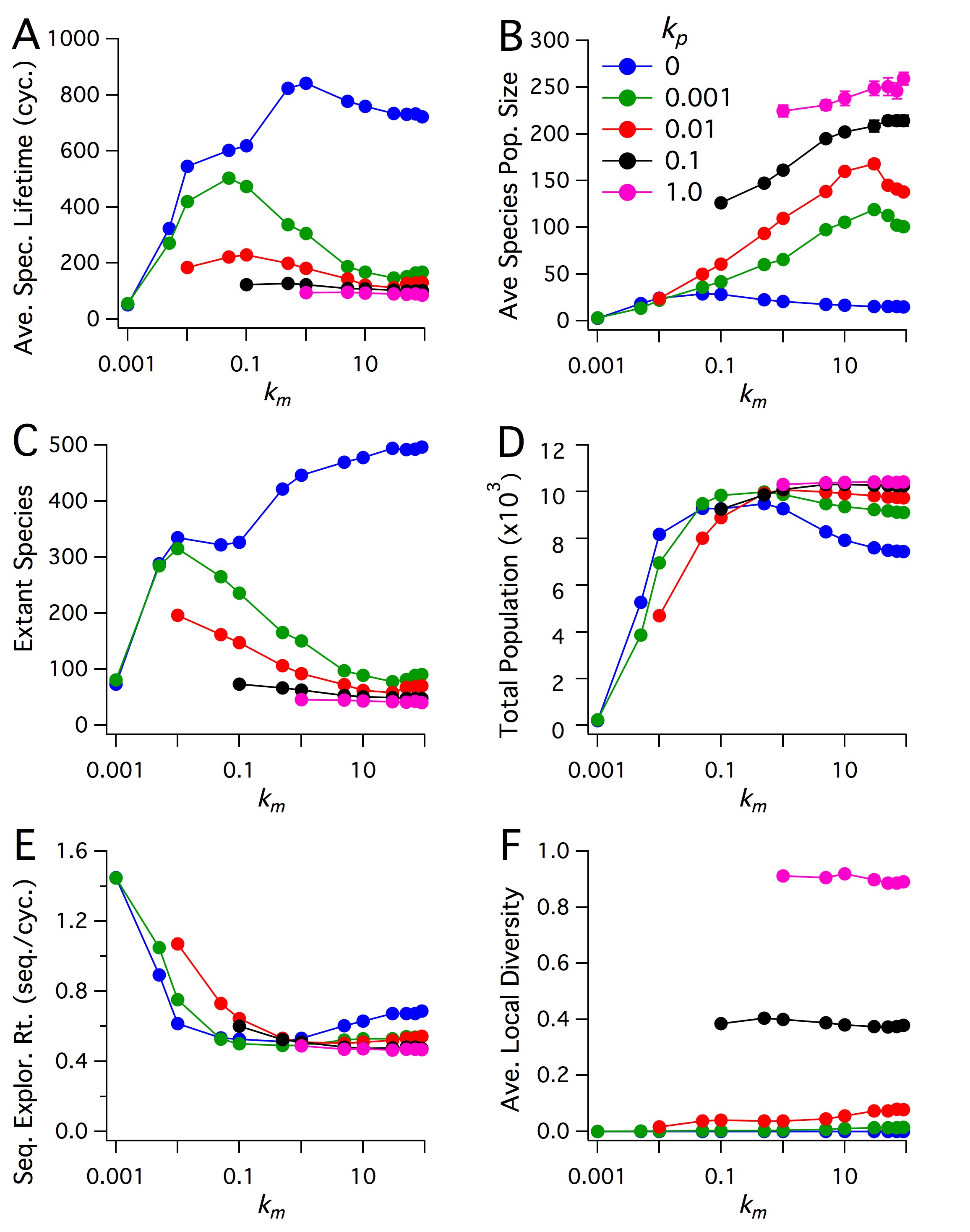}
}
\label{fig:subfigureExample}
\caption[Optional caption for list of figures]{
{\bf Exploring diffusive parameter space.} Plots of quasi steady-state values of Average Species Lifetime (in units of number of cycles), Average Species Population, Extant Species, Total Population Size, Sequence Exploration Rate (in units of number of novel sequences generated per cycle), and Average Local Diversity. Time averages were taken from $t = 2500- 5000$ cycles, and each point is the ensemble average over ten realizations. Error bars denote the sample standard deviation (most are smaller than symbols).  The kinetic rate constants are $k_s = 10^{-7}$, $k_r = 10^{-4}$, and $k_h = 0.1$. The blue data set shows the case where the $k_p = 0$, where polymers are completely immobile. }
\label{fig:D}
\end{figure}

\begin{figure}[ht]
\centering
{
\includegraphics[width=6in]{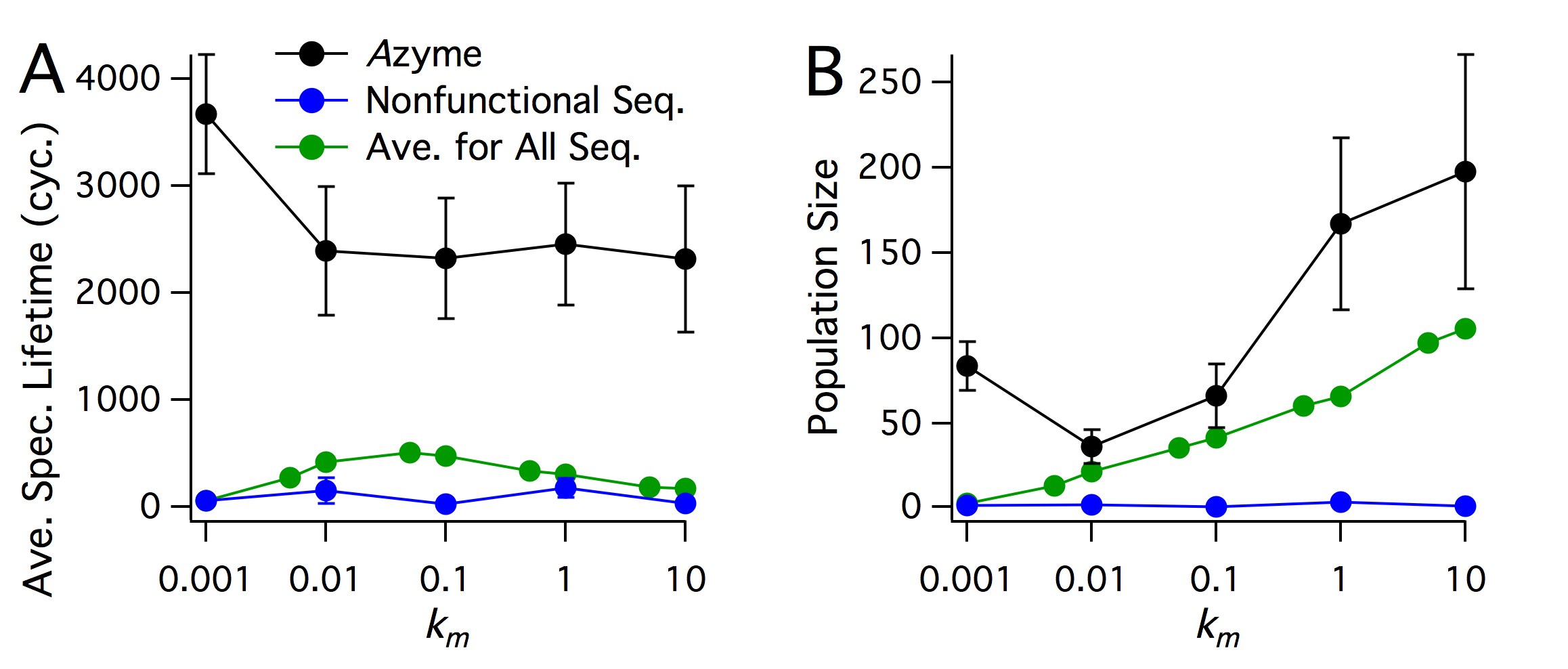}
}
\caption{ 
{\bf Selection for functional sequences.} Plots illustrating the propagation of a functional $A$zyme, compared to nonfunctional sequences. A: Average Species Lifetime (in units of number of cycles), and B: Average Population Size. Each data point is the ensemble average over twenty-five runs, with error bars denoting the sample standard deviation. Kinetic rate constants are $k_s = 10^{-7}$ , $k_r = 10^{-4}$, and $k_h = 0.1$, with a polymer diffusion rate constant of $k_p = 0.001$ sites/cycle. The green points represent overall population statistics for realizations with no $A$zyme (plotted in green in Figure 4). The black points represent statistics for a single functional $A$zyme. The blue points represent statistics for a nonfunctional sequence introduced at the same location and cycle as in the $A$zyme simulations, except with no functionality.} \label{fig:func_runs}
\end{figure}

\begin{figure}[ht]
\centering
{
\includegraphics[width=5in]{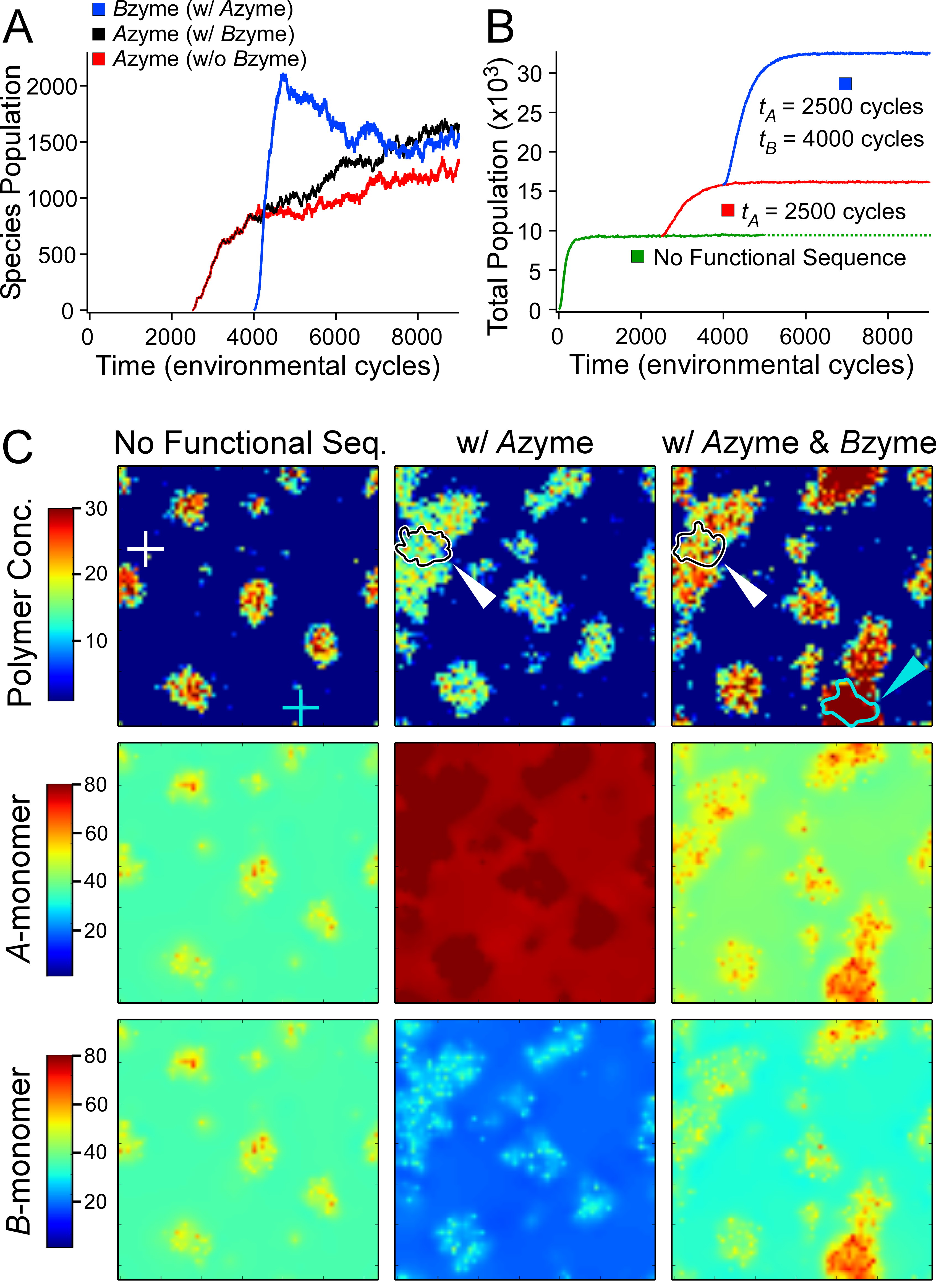}
}
\caption{
{\bf Spatial distribution maps for no functional species, one functional species, and two functional species.} The three scenarios shown are all identical up to $t = 2500$ cycles, at which time the system has achieved a quasi-steady state distribution. In the first scenario, no functional sequences appear. In the second scenario, a functional $A$zyme appears at $t_A$ = 2500. In the third scenario, the same functional $A$zyme appears at $t_A$ = 2500, and a functional $B$zyme also appears at $t_B$ = 4000.  A: Time evolution of the Species Populations of the $A$zyme and $B$zyme. The units of time are in number of cycles. The red curve corresponds to the second scenario, having only the $A$zyme, while the black and blue curves correspond to the third scenario with both enzymes emerging. B: The time evolution of the Total Polymer Population for the three scenarios. C: The spatial distribution of the polymer (total) and monomer concentrations, at $t$ = 5000 cycles. White arrow indicates contour containing $95$\% of $A$zyme polymers, cyan arrow indicates contour containing 95\% of $B$zyme polymers. Kinetic rate constants are $k_s = 10^{-7}$, $k_r = 10^{-4}$, and $k_h = 0.1$, and diffusive rate constants are $k_p = 0.01$ and $k_m = 1.0$ sites/cycle.
} \label{fig:AandBzyme}
\end{figure}

\begin{figure}
\centerline{\includegraphics[width=6.5in]{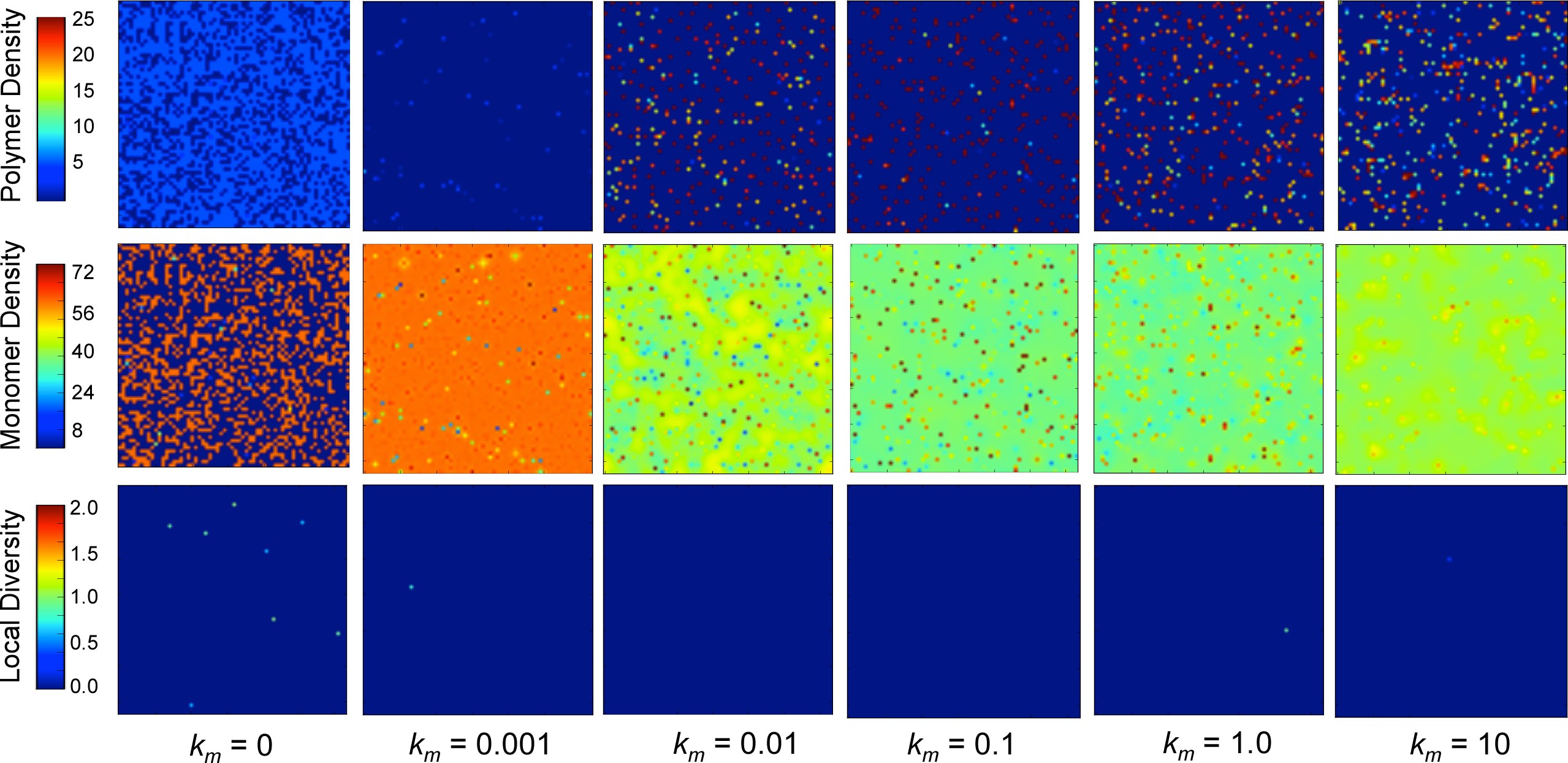}} \label{fig:Dp0maps}
\caption{
{\bf Figure S1. Spatial maps for $k_p = 0$ sites/cycle.} Spatial maps of polymer density (top row), $A = B$ monomer density (middle), and local diversity (bottom) for $k_p = 0$ sites/cycle, with monomer diffusivity increasing from left to right. Polymers do not diffuse and as such are indefinitely stuck on their nucleation site. Snapshots are taken at $t = 2000$ cycles. The kinetic rate constants are $k_s = 10^{-7}$, $k_r = 10^{-4}$, and $k_h = 0.1$. }
\end{figure}

\begin{figure}[!h]
\centerline{\includegraphics[width=5.5in]{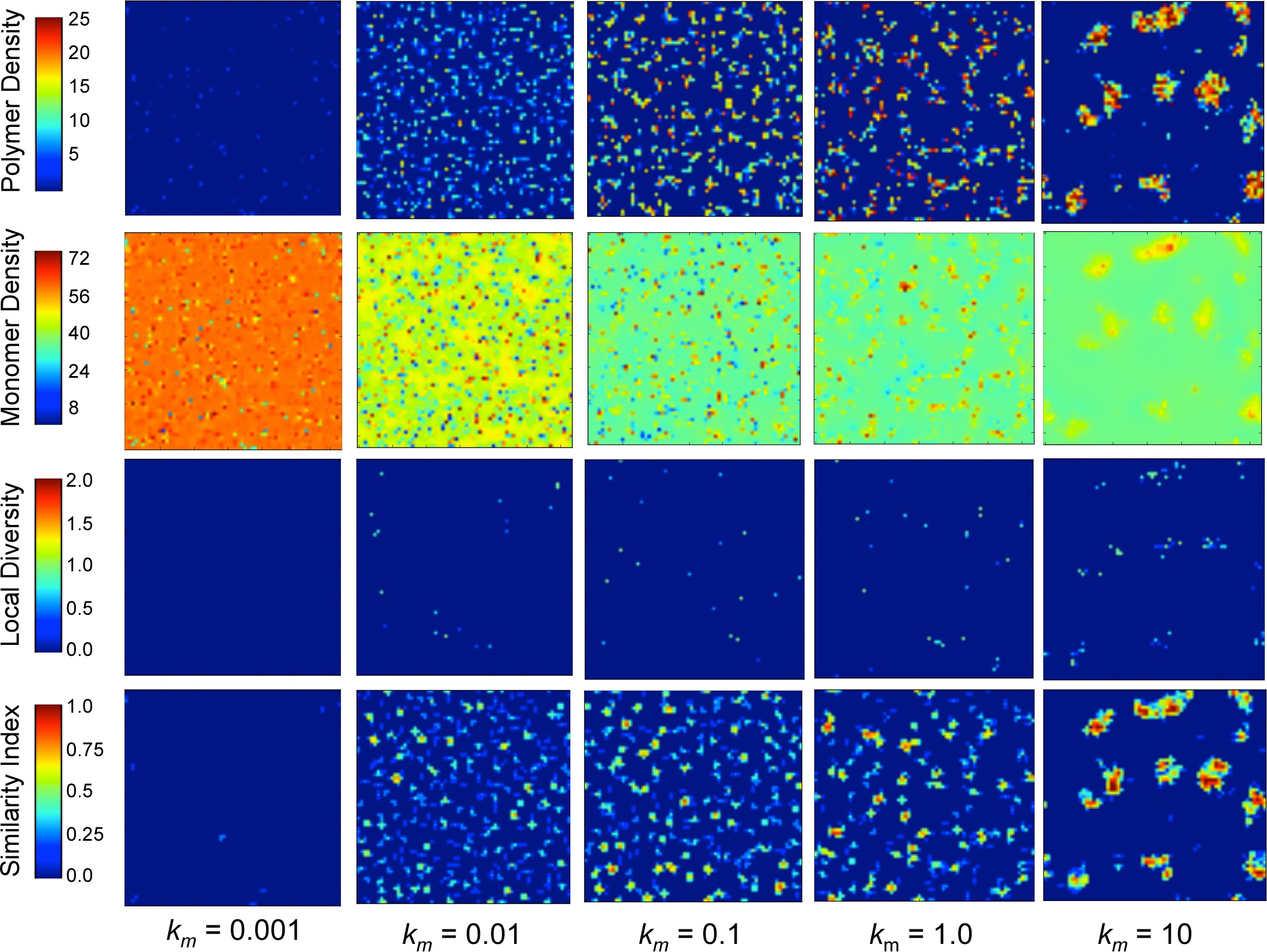}} \label{fig:Dp0001maps}
\caption{{\bf Figure S2. Spatial maps for $k_p = 0.001$ sites/cycle.} Spatial maps of polymer density (top row), $A = B$ monomer density (second row), local diversity (third row), and similarity index (bottom row) for $k_p = 0.001$ sites/cycle, with monomer diffusivity increasing from left to right. Snapshots are taken at $t = 2000$ cycles. The kinetic rate constants are $k_s = 10^{-7}$, $k_r = 10^{-4}$, and $k_h = 0.1$. }
\end{figure}

\begin{figure}[!h]
\centerline{\includegraphics[width=5in]{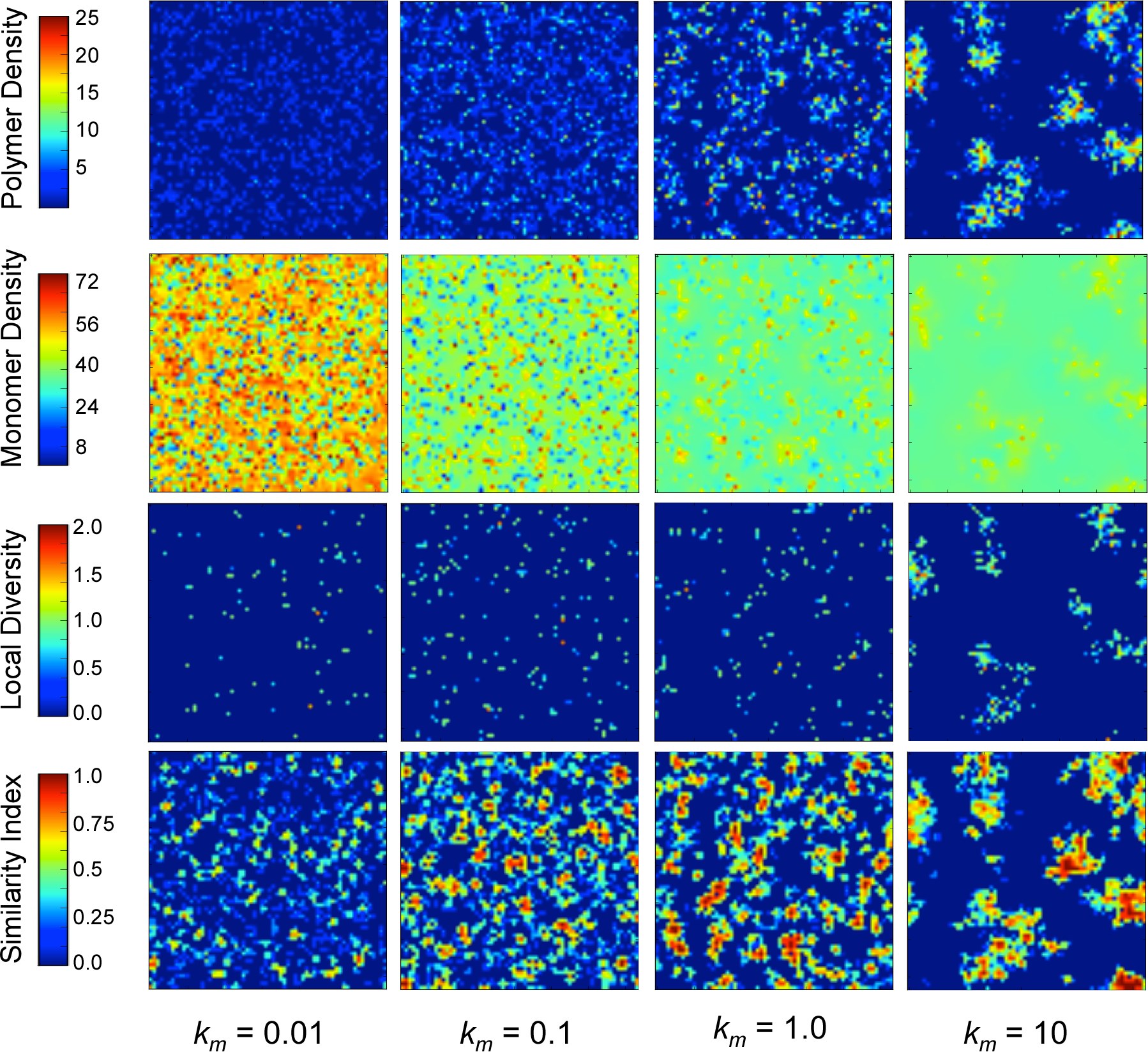}} \label{fig:Dp001maps}
\caption{{\bf Figure S3. Spatial maps for $k_p = 0.01$ sites/cycle.} Spatial maps of polymer density (top row), $A = B$ monomer density (second row), local diversity (third row), and similarity index (bottom row) for $k_p = 0.01$ sites/cycle, with monomer diffusivity increasing from left to right. Snapshots are taken at $t = 2000$ cycles. The kinetic rate constants are $k_s = 10^{-7}$, $k_r = 10^{-4}$, and $k_h = 0.1$. }\end{figure}

\begin{figure}[!h]
\centerline{\includegraphics[width=4.5in]{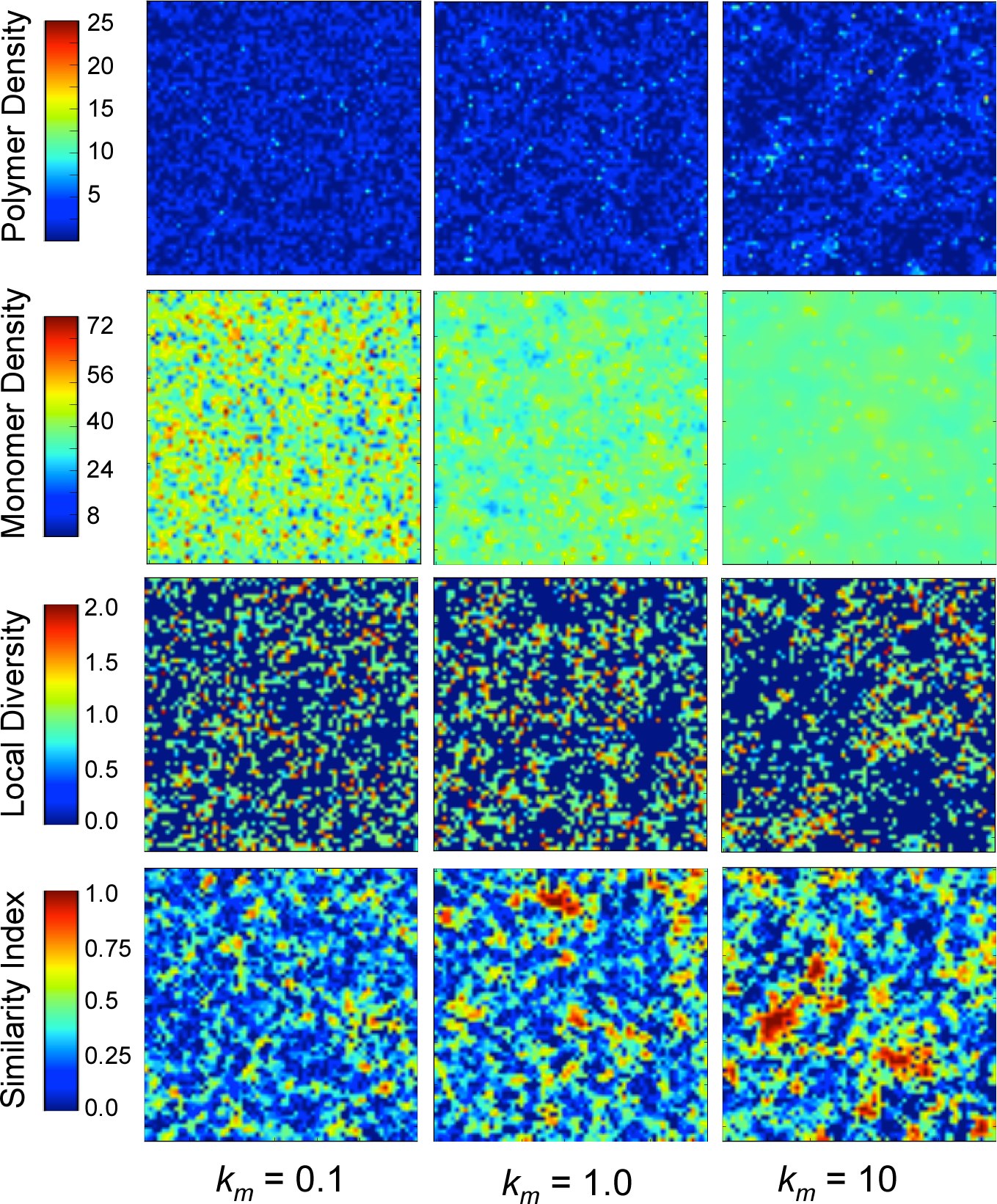}} \label{fig:Dp01maps}
\caption{{\bf Figure S4. Spatial maps for $k_p = 0.1$ sites/cycle.} Spatial maps of polymer density (top row), $A = B$ monomer density (second row), local diversity (third row), and similarity index (bottom row) for $k_p = 0.1$ sites/cycle, with monomer diffusivity increasing from left to right. Snapshots are taken at $t = 2000$ cycles. The kinetic rate constants are $k_s = 10^{-7}$, $k_r = 10^{-4}$, and $k_h = 0.1$. }\end{figure}

\begin{figure}[!h]
\centerline{\includegraphics[width=3.5in]{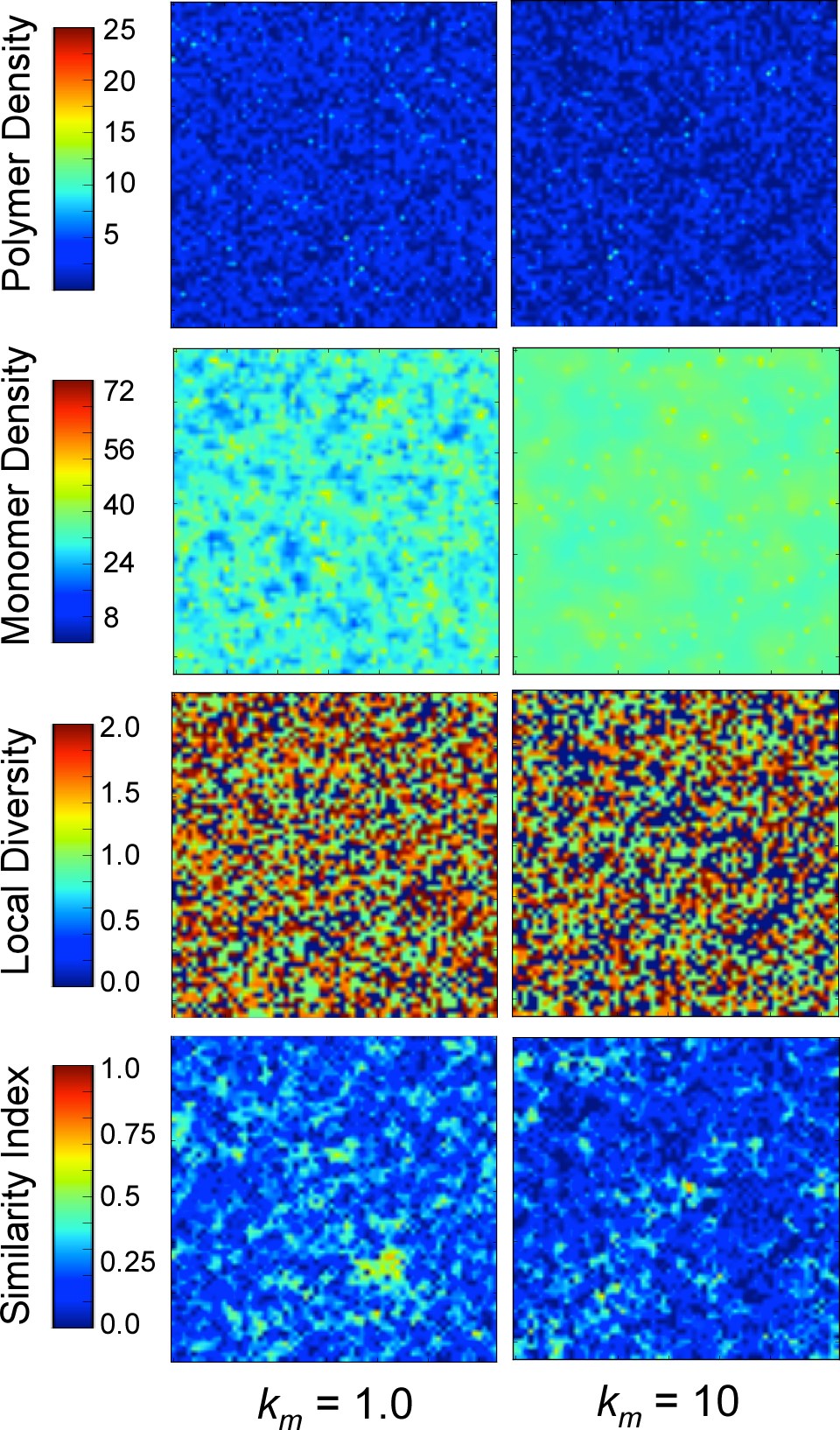}} \label{fig:Dp1maps}
\caption{{\bf Figure S5. Spatial maps for $k_p = 1.0$ sites/cycle.} Spatial maps of polymer density (top), $A = B$ monomer density (second row), local diversity (third row), and similarity index (bottom row) for $k_p = 1.0$ sites/cycle, with monomer diffusivity increasing from left to right. Snapshots are taken at $t = 2000$ cycles. The kinetic rate constants are $k_s = 10^{-7}$, $k_r = 10^{-4}$, and $k_h = 0.1$. }
\end{figure}

\begin{figure}[!h]
\centerline{\includegraphics[width=5in]{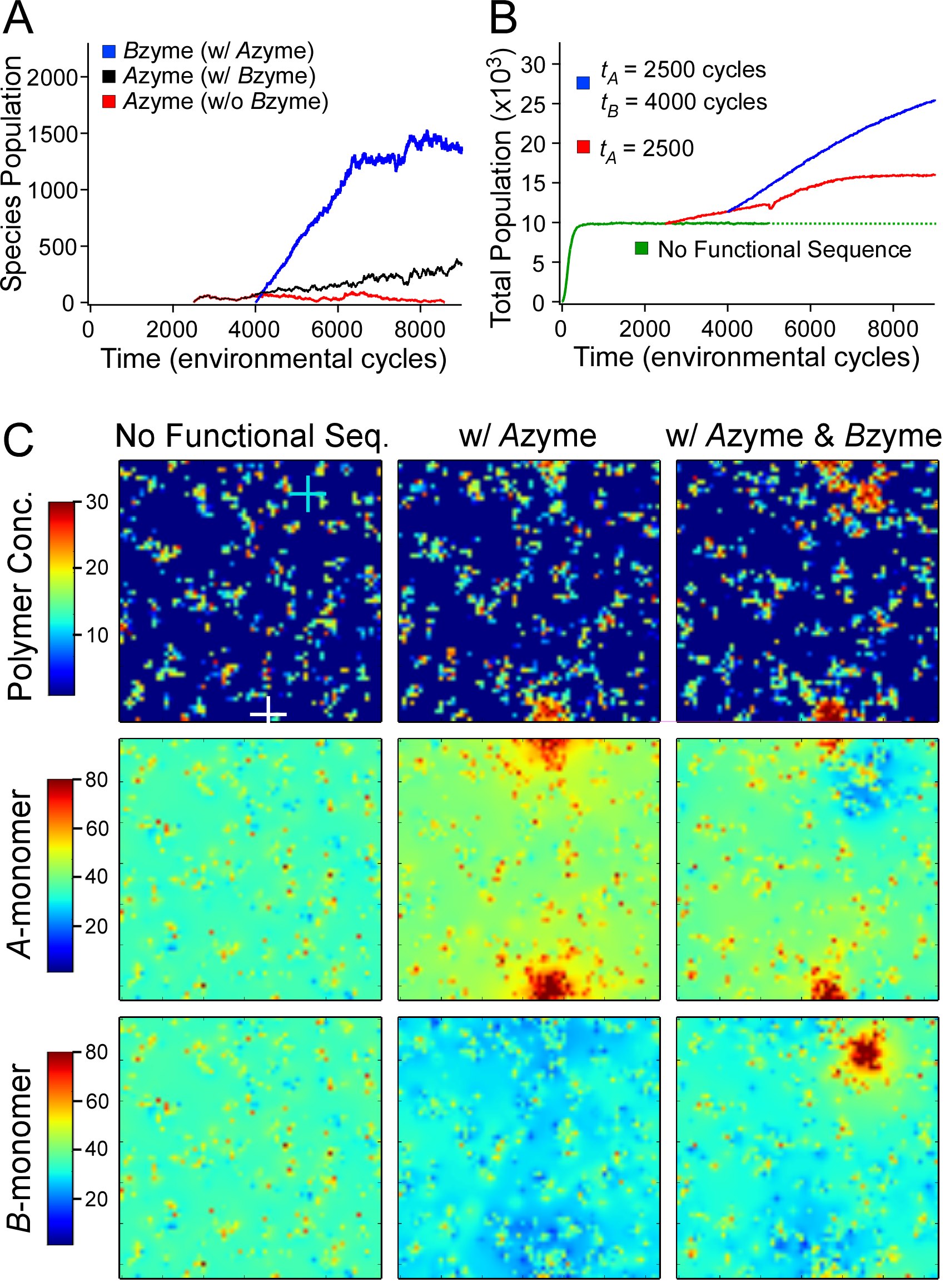}} 
\caption{{\bf Figure S6. Spatial distribution maps for no functional species, one functional species, and two functional species.} The three scenarios shown are all identical up to $t = 2500$ cycles, at which time the system has achieved a quasi-steady state distribution.  In the first scenario, no functional sequences appear.  In the second scenario, a functional $A$zyme appears at $t_A$ = 2500.  In the third scenario, the same functional $A$zyme appears at $t_A$ = 2500 cycles, and the functional $B$zyme appears at $t_B$ = 4000 cycles.  In Panel A, the time evolution of the Species Populations of the $A$zyme and $B$zyme is shown.  The red curve corresponds to the second scenario, having only the $A$zyme, while the black and blue curves correspond to the third scenario with both enzymes emerging.  Panel B shows the time evolution of the Total Polymer Population for the three scenarios. Panel C illustrates the spatial distribution of the polymer (total) and monomer concentrations, at $t$ = 5000 cycles. Kinetic rates are $k_s = 10^{-7}$, $k_r = 10^{-4}$, and $k_h = 0.1$, and diffusive rates of $k_p = 0.001$  and $k_m = 1.0$ sites/cycle.}
\end{figure}

%\end{figure}
%\section*{Tables}
%\begin{table}[!ht]
%\caption{
%\bf{Table title}}
%\begin{tabular}{|c|c|c|}
%table information
%\end{tabular}
%\begin{flushleft}Table caption
%\end{flushleft}
%\label{tab:label}
% \end{table}
\end{document}